\documentclass[final,3p]{elsarticle1}

\usepackage{hyperref}
\usepackage{amsmath,amssymb}
\usepackage{algorithm}
\usepackage{algpseudocode}

\usepackage{graphicx}
\usepackage{hyperref}
\usepackage[caption=false, font=normalsize, labelfont=sf, textfont=sf, farskip=0pt]{subfig}
\usepackage[table,xcdraw]{xcolor}
\usepackage{multirow}
\hypersetup{colorlinks=true}

\journal{XXX}

\begin{document}
\begin{frontmatter}
\title{Singularity Structure Simplification of Hexahedral Mesh
\\via Weighted Ranking}


 \author[label1]{Gang Xu\corref{cor}}
\author[label1]{Ran Ling}
\author[label2]{Yongjie Jessica Zhang}
\author[label1]{Zhoufang Xiao}
\author[label1]{Zhongping Ji}
\author[label3]{Timon Rabczuk}
\address[label1]{School of Computer Science and Technology, Hangzhou
 Dianzi University, Hangzhou 310018, China}
\address[B]{Key Laboratory of Complex Systems Modeling and Simulation,
     Ministry of Education, Hangzhou 310018, China} \address[label2]{Department of Mechanical Engineering, Carnegie Mellon University, USA}
\address[label3]{Institute of Structural Mechanics, Bauhaus-Universität Weimar, Germany}
\cortext[cor]{Corresponding author. Email: gxu@hdu.edu.cn.}

\begin{abstract}
 In this paper, we propose an improved singularity structure simplification method for hexahedral (hex) meshes using a weighted ranking approach. In previous work, the selection of to-be-collapsed base complex sheets/chords is only based on their thickness, which will introduce a few closed-loops and cause an early termination of simplification and a slow convergence rate. In this paper, a new weighted ranking function is proposed by combining the valence prediction function of local singularity structure, shape quality metric of elements and the width of base complex sheets/chords together. Adaptive refinement and local optimization are also introduced to improve the uniformity and aspect ratio of mesh elements. Compared to thickness ranking methods, our weighted ranking approach can yield a simpler singularity structure with fewer base-complex components, while achieving comparable Hausdorff distance ratio and better mesh quality. Comparisons on a hex-mesh dataset are performed to demonstrate the effectiveness of the proposed method.
\end{abstract}


\end{frontmatter}

\section{Introduction}
In recent years, the application of hexahedral (hex) meshes in finite element and isogeometric analysis has become increasingly widespread, because of its good numerical performance, small storage space requirements, and  natural advantage of being able to construct tensor-product splines. However, hex-mesh generation is not yet mature, and it cannot be guaranteed that a good quality initial mesh can be generated in all cases.
For complex shapes and structural models, the octree-based mesh generation method was proposed \cite{zhang2006adaptive,Ito2010Octree}. This method is efficient and robust, and it can ensure a topologically valid and well-formed meshing result. However, it generates a large number of cells and too many singularities. In some scenarios, we do not need a dense mesh and complicated interior structures. Meshes with simple structure and fewer singularities are more conducive to accelerating computational and convergence speed \cite{bourdin2007comparison}. Therefore, it is very important to propose an effective singularity structure simplification method for hex-meshes.

Some research work has contributed to this topic in the past 10 years. In \cite{woodbury2011localized}, an adaptive hex-mesh localization method was proposed. Topological operations such as collapsing and pillowing are used to process the locality, and localized roughening is maintained while maintaining topological connectivity and shape of the input mesh, which provides a basic idea of hex-mesh coarsening. In \cite{gao2015hexahedral}, the mesh structure is simplified according to the reparameterization requirements, and singularity is effectively reduced while maintaining the number of mesh elements. Template matching is used to split patches and eliminate the leading blocks. However, its implementation is very limited and not robust. It cannot simplify self-interleaved and closed loops, resulting in poor results on input meshes obtained from octree-based methods. In \cite{gao2017robust}, a robust hex-mesh structure simplification method was proposed. It is possible that a feasible solution with a simpler and coarser structure exists, but the algorithm might fail to find it. Especially, the ranking method  for the selection of to-be-collapsed base complex sheets/chords is only based on the thickness, and it cannot guarantee to remove most of the singular structures. It will also introduce a few closed-loops and terminate the simplification process in advance. For an initial hex-mesh with many singular vertices, a proper priority ranking algorithm is needed to guide the simplification of the singularity structure. Moreover, a local parameterization is also needed to improve the mesh quality and repair topology structure after simplification.  In this paper, we propose an improved singularity structure simplification method of hex-meshes. The main contribution can be summarized as follows:
\begin{itemize}
\item A new weighted ranking approach for singularity structure simplification is proposed by combining the valence prediction function of local singularity structure, shape quality metric of elements and the width of base complex sheets/chords.
\item A local optimization for SLIM \cite{rabinovich2017scalable} is proposed to improve the uniformity of hex-elements while maintaining the element quality;
\item An adaptive sheet refinement method is proposed to preserve surface features  while maintaining similar number of hex-elements.
\end{itemize}
Based on these improvements, the proposed weighted ranking method can achieve a smaller number of singularities with comparable Hausdorff distance ratio, effectively remove the presence of kinks in the hex-mesh, and yield better mesh quality compared to the thickness ranking method \cite{gao2017robust}.

The remainder of the paper is structured as follows. A review of related hex-mesh generation and mesh simplification is presented in Section  \ref{sec:related}.  Some basic concepts and framework overview are described in Section \ref{sec:overview}. Section \ref{sec:coarsen} presents the sheet and chord collapsing operation of base-complex. The proposed weighted ranking approach is described in Section \ref{sec:weightranking}. Adaptive sheet refinement is presented in Section \ref{sec:refinement}.
 In Section \ref{sec:example}, the experimental results are illustrated. Finally, the paper is concluded and future work is outlined in Section \ref{sec:conclude}.

\section{Related Work}
 \label{sec:related}

\indent\textbf{Hexaheral mesh generation}. Hex mesh has been widely studied for decades. However, an automatic method that can generate high quality hex-meshes for any complex geometry is still unavailable because of the strong topological constraints \cite{shepherd2007topologic}, i.e., the dual chord and the dual sheet. Unlike tetrahedral meshes, any local changes in the mesh would propagate to the whole mesh by dual chords or dual sheets \cite{shepherd2007topologic}, which makes hex-mesh generation a very challenging task. Some methods were devised for specific types of geometries. For example, the mapping method is very preferable for mappable geometries, while the sweeping method \cite{roca2009paving} is often used for swept volumes. By combining with domain partition, they can be applied to complex geometries  \cite{wu2018fuzzy} \cite{roca2009paving}. Based on the idea of paving, several geometric and topological approaches have been proposed for all-hex meshing. Plastering \cite{blacker1993seams} and H-Morph \cite{STEVEN2015H} generate layers of hex elements in geometric ways, whereas the whisker weaving \cite{tautges1996whisker} \cite{ledoux2008extension} method uses spatial twist continuum and generates the topological dual of hex-mesh. Unconstrained plastering \cite{staten2010unconstrained} is extended from plastering.  Different from other paving methods,  it starts from propagating the original geometry boundary instead of a pre-meshed boundary into the interior domain, and hex elements are generated when three propagating fronts intersect each other. The octree-based approach \cite{schneiders1997algorithm} is very robust and can be executed in a highly automatic way, however, it yields poor quality elements near boundary and the final mesh heavily relies on the orientation of the coordinate system. The polycube based meshing approach uses a low distortion mapping between the input model and polycube, and computes the corresponding volumetric mappings. The deformation methods are introduced for polycube construction \cite{Gregson2011All,Liu2015Feature,Hu2016Centroidal,Hu2017Surface}, and frame fields are proposed to guide the polycube construction \cite{Fang2016All,Yu2014Optimizing}.
In \cite{Matthias2011CubeCover}, Nieser \textit{et al.} computes a global parameterization of the volume on the basis of a frame filed to construct hex-meshes. Theoretical conditions on singularities and the gradient frame field are derived for degenerated parameterization, and badly placed singularities can lead to distortion. Based on spherical harmonics representation, Huang \textit{et al.} \cite{Jin2011Boundary} generated a boundary-aligned smooth frame field by minimizing an energy function. Though impressive results were obtained from the frame field based approaches, further efforts are still needed for practical use.

\indent \textbf{Mesh simplification}. Mesh simplification generally reduces the number of elements and maximizes the appearance of the original mesh by performing local coarsening operations. Triangular elements can be combined with the edge flipping operation and local MSL form of the minimum energy function. This method was also applied to hierarchical mesh generation with step by step simplification. In quadrilateral and hex-mesh simplification, similar local operations were also proposed \cite{Tarini2010Practical, shepherd2010adaptive}. Sheets and chords are extracted by the inherent dual structure, and the local operation is simplified for the object \cite{gao2015hexahedral,gao2017robust}.
Recent progress in structure simplification has achieved great success in polycube simplification \cite{Gianmarco2016Polycube} and  hex-mesh optimization \cite{WANG2018103}. In \cite{Gianmarco2016Polycube}, the singularity misalignment problem was solved directly in the polycube space, and the corner optimization strategy was introduced to produce coarser block structured surface and volumetric meshes. Moreover, the induced meshes  are suited for spline fitting. Topology control operations in hex-mesh simplification can also be applied to adjusting low quality mesh elements. In \cite{WANG2018103}, an adjustment strategy for repairing the inverted elements was proposed by combining the basic mesh editing operations with frame field optimization. Based on the singularity structure in the mesh, a base-complex block structure is extracted in \cite{gao2017robust}. Then the simplification operation is performed to collapses base complex sheets and chords while redistributing the distortion based on a volumetric parametrization. However,  the selection of appropriate base complex sheets/chords to be collapsed is only based on their thickness, which will introduce a few closed-loops, cause an early termination of simplification and a slow convergence rate. In this paper, a new weighted ranking function will be proposed by combining the valence prediction function of local singularity structure, shape quality metric of elements and the width of base complex sheets/chords.

\section{Basic concepts and framework overview}
\label{sec:overview}

 The proposed hex-mesh simplification can effectively reduce the singularity structure while maintaining the specified number of elements. In this section, we briefly introduce the definition of singularity structure, base-complex and two types of structure called \emph{base-complex sheet} and \emph{base-complex chord}.

\indent \textbf{Base-complex}. The valence of vertex, edge and face is denoted as  the number of its neighboring hex elements. A vertex is said to be regular if its valence is four on the boundary or eight in the interior. Similar to the regular vertex, an edge is regular when its valence is two on the boundary or four in the interior. Then a series of connected irregular edges with the same valence compose of a singular edge, and its two ending vertices are called \emph{singular vertices}, except the case of closed  singular edges. The singularity structure is composed of these singular edges and singular vertices. According to the above definitions, we can extract the singularity structure of a hex-mesh.
    Each singular edge with a valence of $n$ can be extended to $n$ segmented surfaces, and the valid manifold hex-mesh can be divided into cube-like components by these segmented surfaces (refer to \cite{gao2015hexahedral} for more details). A segmented structure called \emph{base-complex} can be extracted in this way. The  base-complex of the hex-mesh $H$ is denoted as $B = (B_{V},B_{E},B_{F},B_{C})$, where $B_{C}$ is the set of cube-like components (composed of hex elements), $B_{V}$ and $B_{E}$ are the set of 8 corners of each cube-like component and the set of base-complex edges (a series of connected edges between two base-complex vertices) respectively, and $B_{F}$ contains base-complex faces of each component.

\begin{figure}[t]
\setlength{\abovecaptionskip}{0.1cm}
\setlength{\belowcaptionskip}{-0.2cm}
  \centering
    \subfloat[]{
        \includegraphics[width=0.24\linewidth]{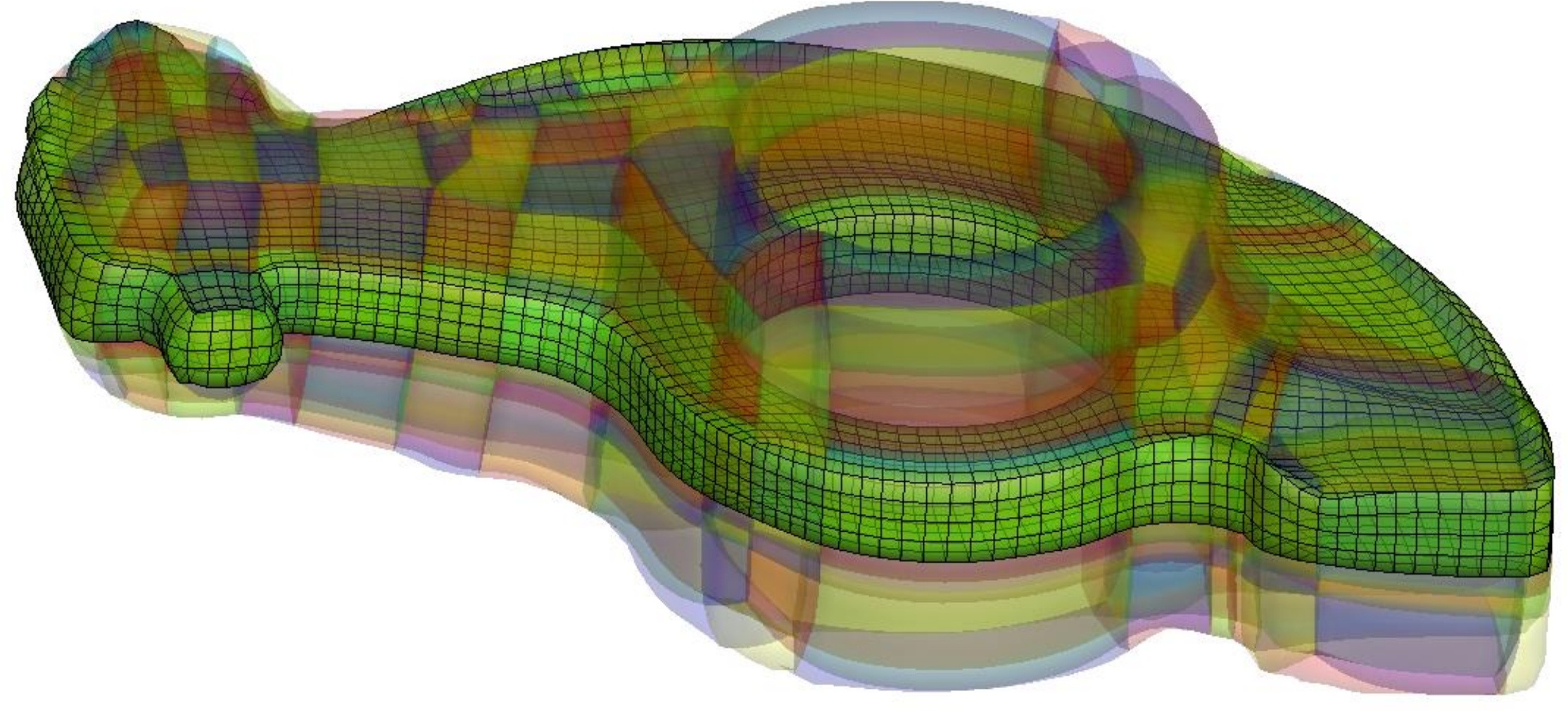}
    }
    \label{1a}
    \subfloat[]{
        \includegraphics[width=0.25\linewidth]{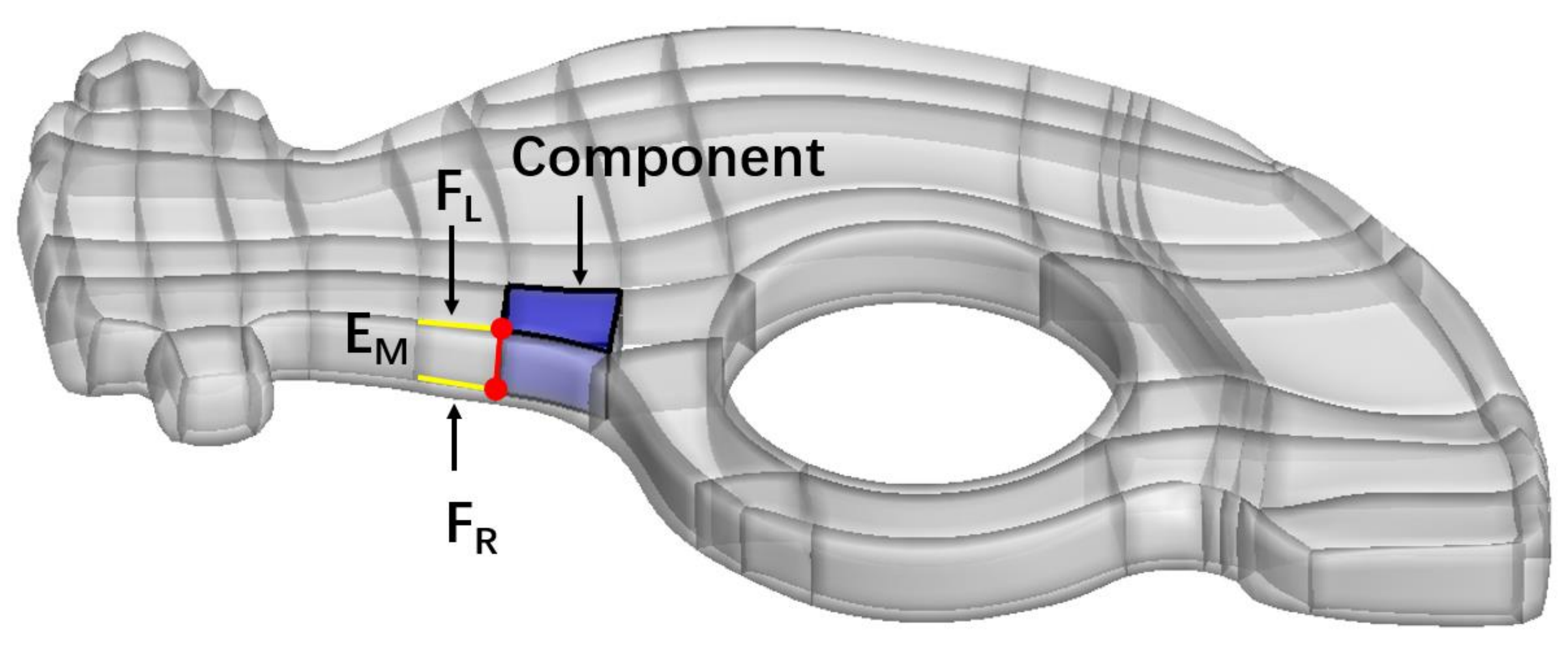}
    }
    \label{1b}
     \subfloat[]{
        \includegraphics[width=0.25\linewidth]{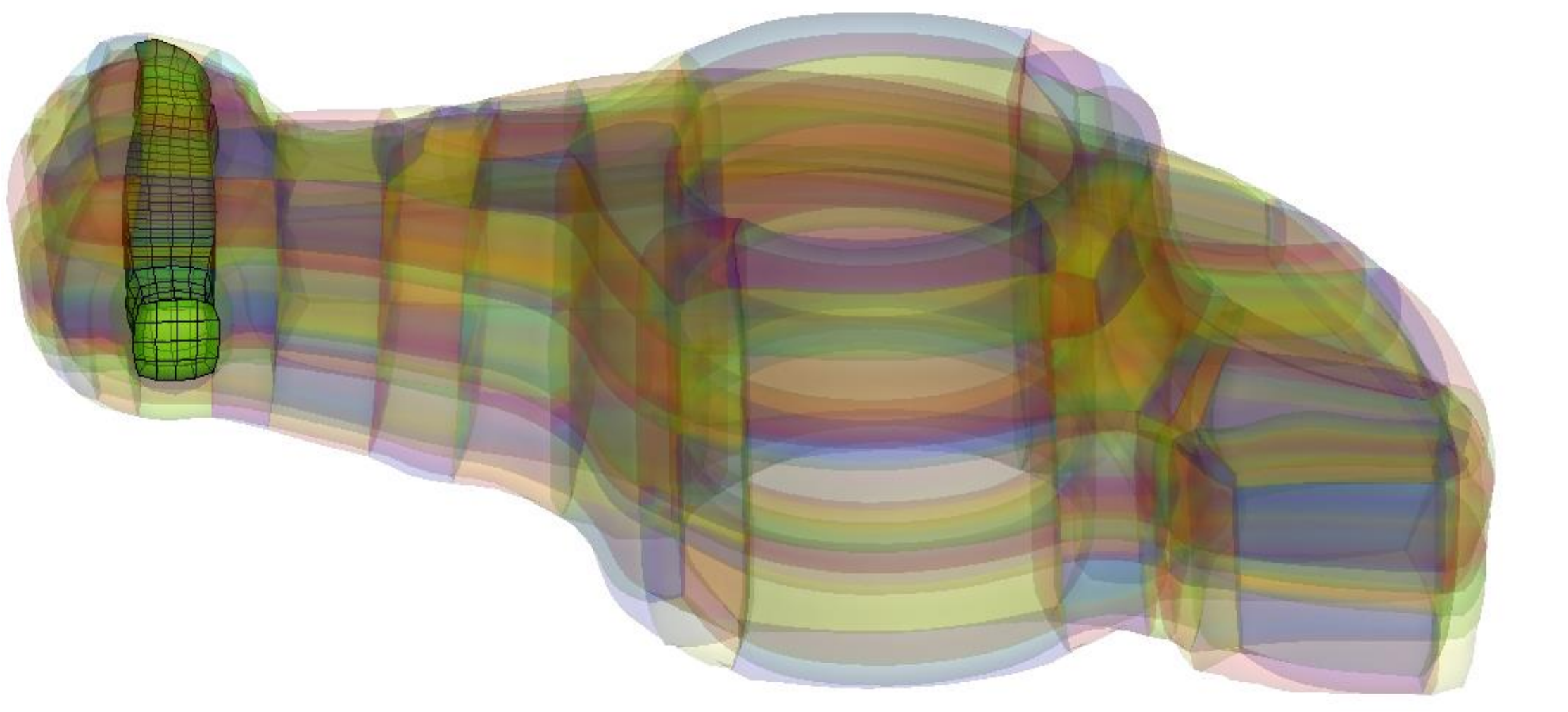}
    }
    \label{1c}
     \subfloat[]{
        \includegraphics[width=0.15\linewidth]{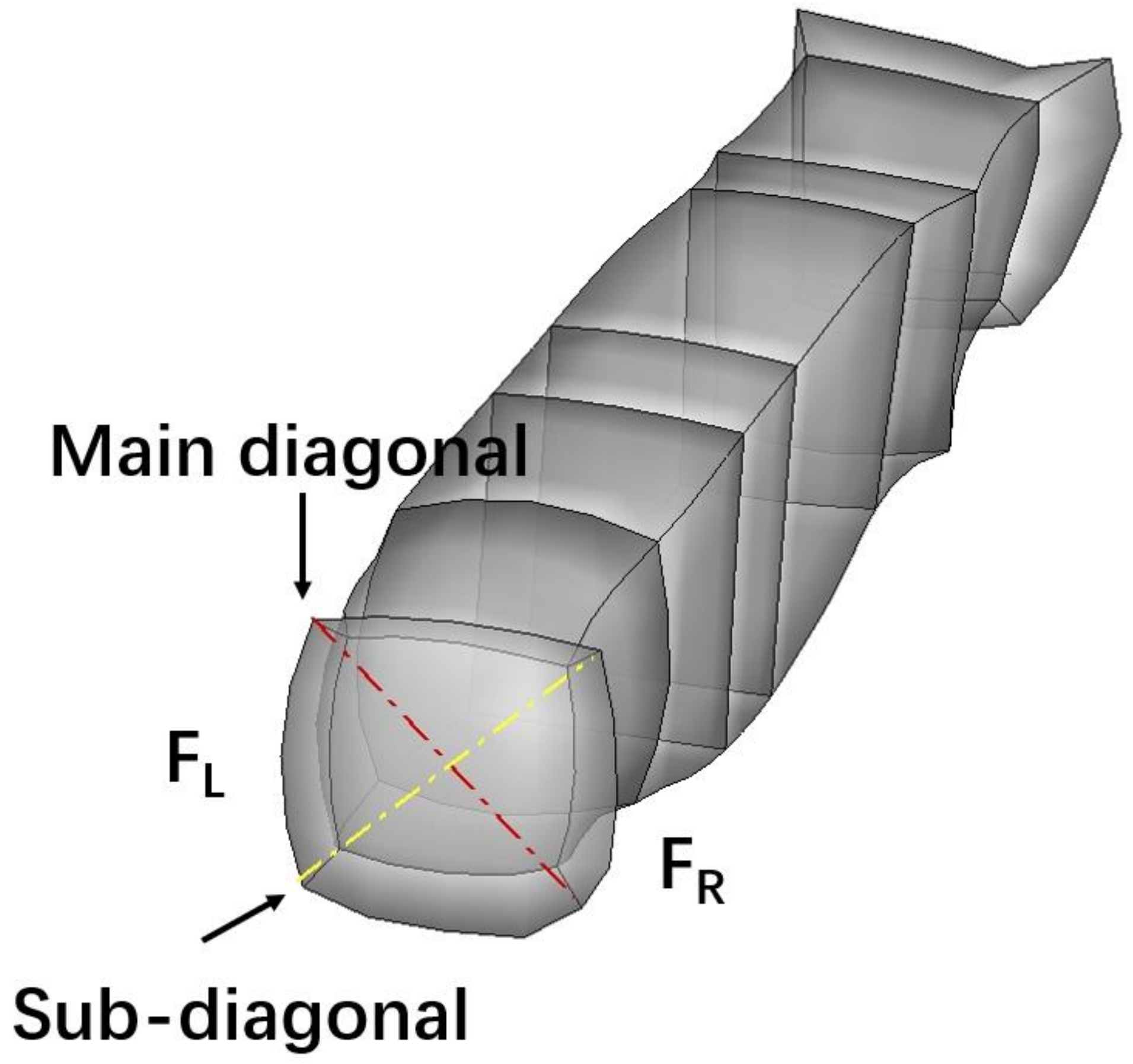}
    }
    \label{1d}
    \caption{(a) The base-complex sheet (green elements) consists of the left surface $F_L$, the right surface $F_R$ and the middle volume $E_M$, with the edge pair (yellow edges) and the vertex pair (red dots) shown in (b). (c) The green elements form a base-complex chord, where $F_L$ and $F_R$ in (d) can be determined from the main diagonal direction.}
    \label{fig1}
\end{figure}

Base-complex sheet and base-complex chord can be extracted based on the base-complex  structure. Since each of these components aligns with its adjacent components with $C^0$ continuity, and the singularities are located at its eight corners
and three groups of four topologically parallel base-complex edges. Removing
components can effectively simplify singularity structure by collapsing base-complex sheets and chords. The base-complex sheet $S$ consists of three parts: the left surface $F_{L}$ (or the right surface $F_R$) contains all base-complex vertices, edges and faces in the boundary of the left (or right) part, and the middle volume $E_{M}$
contains the base-complex edges with two end nodes on $F_{L}$ and $F_{R}$
respectively. Topology elements in $F_{L}$ and $F_{R}$ can form
element groups. Base-complex chord has a similar definition, in which two sides follow the main diagonal direction.  Fig. \ref{fig1} shows the structure of base-complex sheet and base-complex chord.

\begin{algorithm}[t]
  \caption{Framework of singularity structure simplification}
  \label{alg:Framwork}
  \begin{algorithmic}[1]
    \Require
      A hex-mesh $M$;
      Target number of mesh elements, $N_{c}$;
      Target reduction ratio of components, $N_{s}$;
      The current number of elements, $n_{s}$;
    \Ensure
      Hex mesh with a simplified base-complex, $m_{out}$;
    \State Extract the base-complex structure $B = (B_{V},B_{E},B_{F},B_{C})$ from $M$, secondary detecting until no irregular components is found;\label{code:fram:extract}
    \State Extract all base-complex sheets and chords which satisfy filtering criteria, then push them into two priority queues $S_{sheet}$ and $S_{chord}$ separately, the queue lengths are $k_{s}$, $k_{c}$;
    \label{code:fram:trainbase}
    \State Find the top-ranked base-complex sheet and $\biggl\lfloor k_{c}/k_{s} \biggr\rfloor$(value takes 3 when greater than 3) base-complex chord to remove, when $n_{s}<N_{s}$,
    go to \emph{Step 5};
    \label{code:fram:add}
    \State Remove the sheet/chord using the local parametrization , and use local regularization smoothing for the local step in the framework. If a valid mapping parameterization is not found or the quality metric is below the threshold, remove the next sheet/chord until a successful operation is performed. Otherwise, go to \emph{Step 5}, an adaptive refinement will be performed when the Hausdorff distance ratio goes up to the user-specified threshold $r_{h}$;
    \label{code:fram:classify}
    \State If the specified threshold $N_{s}$ is not satisfied, go back to \emph{Step 1}, and when the number of elements is smaller than $N_{c}$, perform adaptive refinement;
    \State After finishing the simplification process, perform a global optimization operation, return $m_{out}$.
  \end{algorithmic}
  \label{algorithm1}
\end{algorithm}

\indent \textbf{Framework overview}. As shown in Algorithm \ref{algorithm1}, we propose an improved singularity structure simplification method for hex-meshes while maintaining the shape boundary and the target number of elements. After comparison with experimental data, we find that the collapsing order of base-complex sheets and chords has a significant effect on the final simplification results. Hence, we propose an optimized weighted ranking approach for components removing based on the analysis of edge valence. All the base-complex sheets/chords are ranked with the valence error by minimizing an objective function of singularity structure. With the proposed method, the singularity structure complexity of a hex-mesh  decreases rapidly. Furthermore, a few close-loops and entangled sheets can be commendably eliminated, leading to a high simplification rate. In addition, two extral ranking terms are adopted to maintain the elements quality and shape boundary.
In the simplification, sheet refinement is performed
to obtain a similar number of elements as the target number $N_{s}$. We propose an adaptive sheet refinement method based on the point-sampled Hausdorff distance on surface, which can improve the hex-element uniformity and reduce the error between the input and output hex-mesh geometry. To locally improve the uniformity and aspect ratio, we also propose a local regularization optimization in the parametrization for sheet/chord collapsing.

\section{Coarsening operators on hex-meshes}
\label{sec:coarsen}
In this section, we introduce two local coarsening operations on hex-meshes: the base-complex sheet collapsing operation and the base-complex chord collapsing operation, which are two generalized concepts to reduce singularity structure complexity of hex-meshes. The base-complex sheet collapsing operation is mainly applied to change singularities globally, and has a bigger impact on the boundary shape. The base-complex chord collapsing operation is used to optimize local singularity structure, especially for removing edge pairs with a valence of 3$\sim$5. These two operations may introduce non-manifold and doublet configurations as shown in Fig. \ref{fig2}. Moreover, the collapsing operations may lead to local higher complexity which should be prevented. Hence, several filtering criteria will be proposed to avoid these problematic cases.

\subsection{Base-complex sheet collapsing operation}

\begin{figure}[!htb]
\setlength{\abovecaptionskip}{0.0cm}
\setlength{\belowcaptionskip}{-0.2cm}
  \centering
        \includegraphics[width=1.0\linewidth]{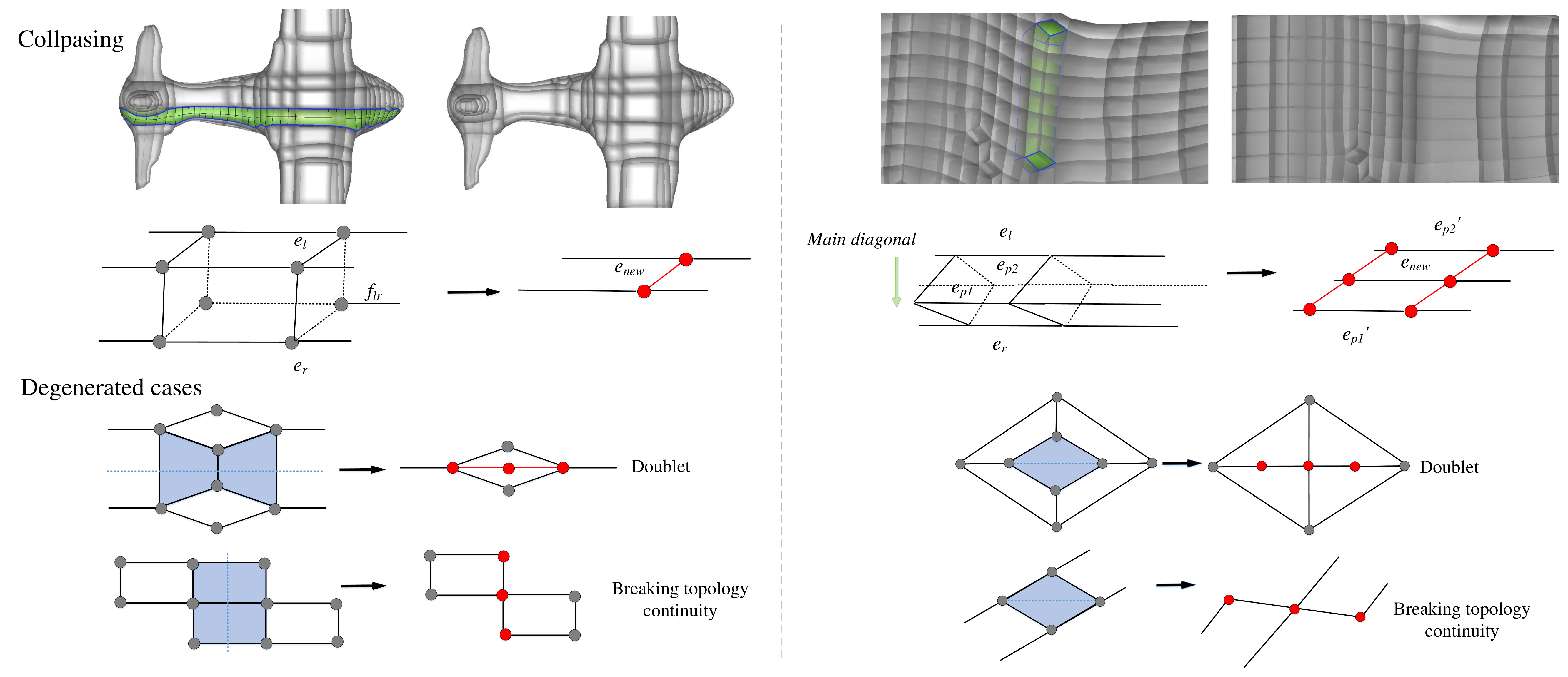}
    \caption{ Left: Base-complex sheet collapsing operation and 2D degenerated cases; Right: Base-complex chord collapsing operation and 2D degenerated cases. The red components may change the edge valence.} \label{fig2}
\end{figure}

 A base-complex sheet collapsing operation similar to \cite{gao2017robust} will be adopted here. Both sides of a sheet can be found by components, and then we remove the middle part of the base-complex sheet and preserve the side of $F_{L}$ or $F_{R}$. Finally,  parametrization is employed to relocate these vertices within the $\beta$-ring neighborhood region ($\beta$ is set to 4 as \cite{gao2017robust}). Before sheet collapsing, several filtering criteria are used to detect whether it should be put into the priority queue.

\indent\textbf{Valence prediction}. Edge pairs in $F_{L}$ and $F_{R}$ are collapsed into a single edge, and the corresponding edge valence may be changed. Generally, the valence of an inner edge is greater than two; otherwise, the adjacent elements will degenerate or form a doublet configuration (two hexahedra share two or more faces as in Fig. \ref{fig2}), which is forbidden in our framework. For the edge pair of $e_{l}$ and $e_{r}$ in a non-self-intersection sheet, if the new edge is denoted as $e_{n}$, then the valence of $e_{n}$ can be computed as follows:
\begin{equation}
S_{after}(e_{l},e_{r})=\left\{
\begin{array}{rcl}
v(e_{l})+v(e_{r})-2, &   &{P(e_{l},e_{r})=0}\\
v(e_{l})+v(e_{r})-4, &   &{P(e_{l},e_{r})=1}
\end{array} \right.
\end{equation}
where $v(e)$ is the valence of a base-complex edge. For the base-complex face directly connects $e_{l}$ and $e_{r}$, $P(e_{l},e_{r})=1$ when it is on the boundary; otherwise, $P(e_{l},e_{r})=0$. The base-complex face is either on the boundary or in the interior of hex-mesh.

\indent\textbf{Boundary shape}. The feature vertices/lines are extracted in the initialization stage, and in order to preserve sharp features, the sheet and chord contain sharp feature vertices are not allowed to be removed.  Moreover, the base-complex sheet is not collapsed when the feature edges lie on base-complex edges. In the collapsing operation, we use a similar way for hex-mesh sheet collapsing. Firstly, we find all elements for both sides, then choose the temporary positions for vertex pairs.
The topology element pairs in $F_{L}$ or $F_{R}$ are only preserved in one side, then we remove all hexahedra between these two sides. In the optimization step, local parameterization \cite{gao2017robust} is adopted. The boundary shape error and interior distortion will be distributed to $\beta$-ring neighboring elements by solving $\min \limits_{V}E(V)$ with the SLIM approach \cite{rabinovich2017scalable}.

\subsection{Base-complex chord collapsing operation}

The base-complex chord collapsing operation is mainly used to optimize bad singularity structure locally. It only has effect on one column of base-complex components. Different from chord collapsing in hex-mesh that merging four vertices per group into a new position, Fig. \ref{fig2} shows the 2D case of chord collapsing. We extract two pairs of opposite base-complex edges, and merge them along the diagonal direction. Here the collapsing direction is denoted as the main diagonal direction and the orthogonal direction along boundary is referred as the sub-diagonal direction. If the number of elements in opposite edges is different, we will collapse several sub-sheets before applying the collapsing.

\indent\textbf{Collapsing direction}. The collapsing direction can be chosen in two directions, and the collapse following these two directions will have quite different influence on singularity structure. The valences of base-complex edges in two sides along the main diagonal direction may be changed. Here, we only consider the four groups of topology-parallel base-complex edges in the surface of chord following the direction of dual string. We compute the predicting valence of these created base-complex edges, and obtain the valence difference between the created edge and the regular edge. Our objective is to remove pairs with a valence of 3$\sim$5 without introducing  high  valence singularities. In this paper, we measure the difference between the predicted valence and the regular valence using
\begin{equation}
D_{v}(c) = \sum\limits_{i=1}^{k}(\left| v(e_{p1i})-p(e_{p1i})-1 \right| + \left| v(e_{p2i})-p(e_{p2i})-1 \right| +\left| v(e_{li})+v(e_{ri})-\min(p(e_{li}),p(e_{ri}))-2 \right|),
\end{equation}
$$
D(c) = \min(D_{v1}(c),D_{v2}(c)), \quad\quad
p(e) = \left\{
\begin{array}{rcl}
3,   &   &e\in E_{surface}\\
4,   &   &e\in E_{inner}
\end{array}\right.
$$
where $e_{p1i}$ and $e_{p2i}$ are base-complex edges in the sub-diagonal direction, $e_{li}$ and
$e_{ri}$ are in the main diagonal direction as shown in Fig. \ref{fig2}, $k$ is the number of contained components of the base-complex chord. We choose the optimal collapsing direction by minimizing $D(c)$. In our experiments, the chord collapsing operation is not allowed when $D(c)/3k>0.9$.
In addition, we implement an easy-to-detect method in advance to improve efficiency. The four groups of parallel edges containing less than 2 groups are all singularities, which will not remove the singular edges locally while collapsing. This kind of chord will not be pushed to the priority queues.

The above operations are iteratively performed during simplification. Base-complex sheet collapsing can make significant impact on mesh globally, but it is extremely difficult to remove self-intersection sheets with complex tangles and close-loop configurations without creating vertices with high valence. Base-complex chord collapsing is used to eliminate the entangled regions, and it contributes to improving the simplification ratio of sheets.  Experimental results show that a higher simplification rate can be achieved by alternately performing these two operations.

\begin{figure}[htb]
\setlength{\abovecaptionskip}{0.1cm}
\setlength{\belowcaptionskip}{-0.2cm}
  \centering
  \subfloat[]{
        \includegraphics[width=0.52\linewidth]{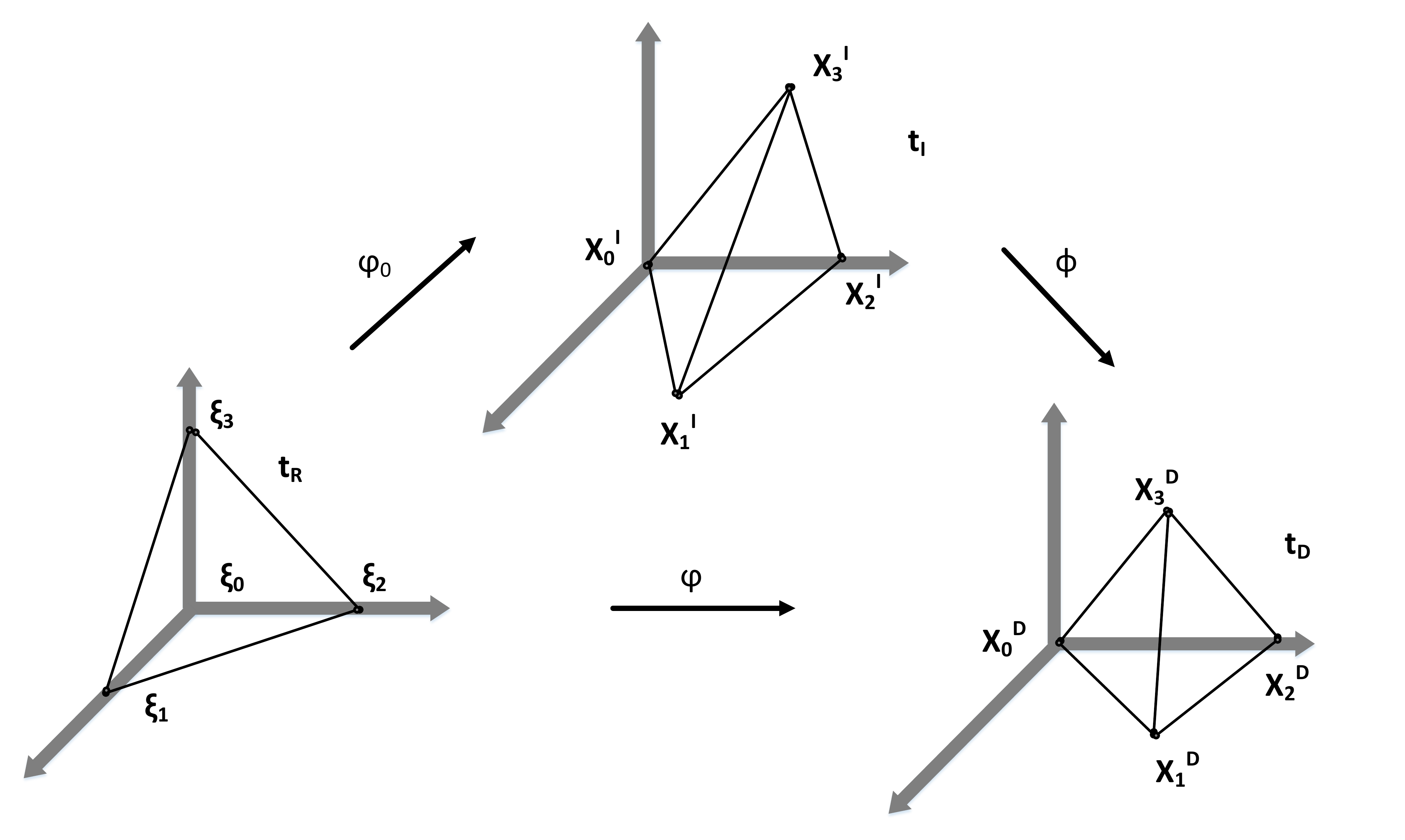}}
    \label{3a}
   \subfloat[]{
        \includegraphics[width=0.39\linewidth]{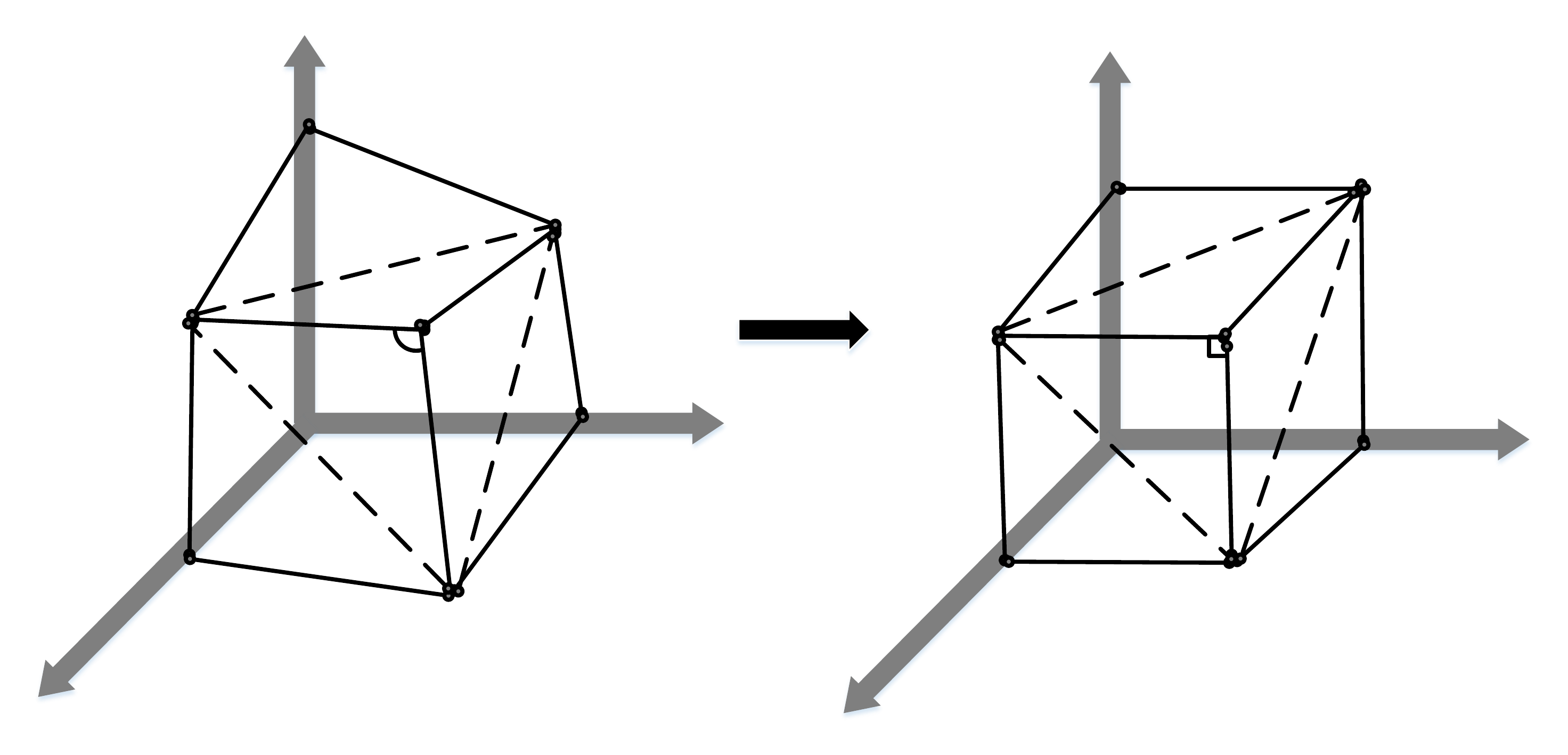}}
    \label{3b}
 \caption{(a) The mapping from a reference tetrahedron (left) to the origin shape (middle) and deformed shape (right). (b) The mapping from a hex element (left) to the element with ideal shape (right) of five terahedra.} \label{fig3}
\end{figure}

\subsection{Local parameterization for uniformity improvement}

After collapsing arbitrary sheets/chords, we apply a local parametrization \cite{gao2017robust} based on SLIM  \cite{rabinovich2017scalable} to relocate points within the collapsing region. The framework of SLIM uses the local/global algorithm \cite{Gotsman2010A}, and solves the distortion term globally while fixing the rotation as computed in the local step. In 3D case, the mapping from the original tetrahedral element to a deformed shape in a local orthogonal frame can be denoted as a Jacobian, and the deformation can be expressed indirectly by a transformation from the tetrahedron with three orthogonal edges to both shapes as shown in Fig. \ref{fig3}. The mapping between the reference element to the original element is defined as
\begin{equation}
        \varphi_0 : t_R\rightarrow t_I , \quad x^I = W^I\xi + x^I_0,
\end{equation}
     where
\begin{equation}
    \begin{split}
        W^I= (X^I_1-X^I_0\quad X^I_2-X^I_0\quad X^I_3-X^I_0)=\left(
        \begin{matrix}
            x^I_1-x^I_0& x^I_2-x^I_0 & x^I_3-x^I_0\\
            y^I_1-y^I_0& y^I_2-y^I_0 & y^I_3-y^I_0\\
            z^I_1-z^I_0& z^I_2-z^I_0 & z^I_3-z^I_0
        \end{matrix}\right)
    \end{split}
\end{equation}
     is a constant matrix. Similarly, the mapping between the reference element and the deformed element is:
    \begin{equation}
        \varphi : t_R\rightarrow t_D,\quad x^D = W^D\xi + x^D_0.
    \end{equation}
Since $W_D$ and $W_I$ are affine matrices, finally the Jacobian $\phi$ of $t_I\rightarrow t_D$ can be denoted as:
     \begin{equation}
         \phi = \varphi \circ \varphi_0^{-1}.
     \end{equation}
Our experiments show that adjusting the Jacobian of a transformation to the target shape in a local operation can lead to an ideal mesh result after global simplification. In this paper, we also propose a local optimization strategy to move vertices within the collapsing region during parameterization. For edges in the collapsing region, their length will be re-scaled while maintaining the element quality.

    Let $M = (V,K)$ be the mesh of the parameterized region,   $V$ is the set of nodes, and $K$ is the set of connectivity information, including nodes $\{i\}$ and edges $\{i,j\}$. The discrete operator on $M$ is defined as
    \begin{equation}
        (Lv)_i = \sum_{j}\omega_{ij}(v_i - v_j),
    \end{equation}
  and the iterative form can be defined as
  \begin{equation}
    \begin{split}
       v_i^k = \sum_{j=1}^{N_i} \ w_{ij}v_j/N_i, \quad
        w_{ij} = \omega_j/\sum_{j=1}^{N_i}\omega_{ij},\quad
\omega_{ij} = \left\{
\begin{array}{rcl}
1,    &   &v_i\in V_{in} \\
0,     &   &v_i\in V_{bdy} \\
0,       &   &v_i\in F_{L} \cup F_{R}
\end{array} \right.
    \end{split}
    \end{equation}
and the iteration is terminated when the threshold of variance $\epsilon$ is reached,
    \begin{equation}
        \frac{\left[ \sum_i(v_i^k-v_i^{k-1})^2 \right]^{1/2} }{\left[ \sum_i\left[ (x_i^{k-1})^2+(y_i^{k-1})^2+(z_i^{k-1})^2 \right] \right]^{1/2}} < \epsilon
    \end{equation}
where $i$ and $j$ are the vertex labels, $N_j$ is the number of neighboring vertices of the $j$th vertex, $V_{in}$ is the set of inner vertices (not including vertices in $F_L$ and $F_R$), and $V_{bdy}$ is the set of vertices on the boundary.


\section{Weighted ranking for structure simplification}
\label{sec:weightranking}

Many hex-mesh generation method such as octree-based and frame-field methods often yield unnecessary interior singularities. The resulting hex-mesh will have a large number of small components in base-complex since the singular edges are distributed along the twelve edges of a cube-like component. The number of singularities can be progressively decreased by performing collapsing operations based on components, and the simplified singularity structure is obviously different with various collapsing sequences. In this paper, we introduce a weight ranking sequence, which can choose the optimal candidate to remove iteratively. The ranking sequence aims to remove singularities within fewer iterative steps. We formulate this problem as an energy minimization framework, and introduce a \emph{valence  term} related to the valence difference caused by collapsing to achieve a rapid removal of singularities. In addition, optimization will be performed after each simplification step, and the distortion error caused by collapsing is distributed to neighboring elements and sheets. On the other hand,  the collapsing operation is also under the constraints that the resulting elements should not be inverted and the max Hausdorff distance ratio $r_{h}$ should be kept. Hence, the sheet/chord removal leading to less mesh distortion will have the collapsing priority. From this motivation, we also introduce two extra ranking terms, called the \emph{distortion term} and the \emph{width term}. In our framework, the ranking function is a  combination of the valence term, the distortion term and the width term, which is more robust than the previous ranking method \cite{gao2017robust} only based on the thickness of base-complex sheets/chords.

\subsection{Ranking method of base-complex sheet}

In the base-complex sheet ranking sequence, we combine the \emph{valence term}, the \emph{distortion term} and the \emph{width term} as the normalized form
\cite{daniels2008quadrilateral}. The ranking function which can greatly improve the simplification rate of base-complex components is defined as
\begin{equation}
E_s(s) = k_{sq}(1-e^{-E_{sq}(s)}) + k_{sd}(1-e^{-E_{sd}(s)})
+ k_{sv}(1-e^{-E_{sv}(s)})
\end{equation}
where $k_{sv}$, $k_{sd}$ and $k_{sd}$ are weights of different ranking terms.
In our implementation, the valence term $E_{sd}(s)$ has the biggest weight, i.e, $k_{sv} = 0.4$, $k_{sd} = 0.6$ and $k_{sd} = 0.2$. We also control the value of each term within $(0,2)$ to reduce the impact of the actual numerical size.

\indent\textbf{Valence term}. The proposed weighted ranking algorithm for base-complex sheet collaping mainly focuses on the valence difference of singular edges during the simplification.
It has been proved in \cite{gao2017robust} that the singularities of a hex-mesh will be progressively simplified within a finite number of iterations, and the number of components will decrease while reducing the valence of singular edges. In this paper, we propose an indirect energy function of valence difference between the current mesh and the mesh without singularities. For the mesh with singular base-complex
edges set $S(e) =\left\{ e \mid e \in B_{E} \mid \textit{e}
 \text{ is  singular} \right\} $, the energy function is defined as
\begin{equation}
 E(m) = \sum\limits_{e \in S(e)} \left| v(e)-p(e) \right|.
\end{equation}
Since the simplification process is  based on two kinds of collapsing operations, and the singular edges are only located in $F_{L}$, $F_{R}$ and $E_{M}$, then the energy function $E(m)$ has a local representation on the base-complex sheet when it is  collapsed,
\begin{equation}
 E(m) = \sum\limits_{i = 0}^{n} (  -\sum\limits_{e_{m} \in E_{M}} \left| v(e_{m})-p(e_{m}) \right|
 + \gamma \sum\limits_{e_{lr} \in F_{L},F_{R}} \left| v(e_{lr}^{\prime})-p(e_{lr}) \right| ) , \quad  \gamma \in (0,0.5]
\end{equation}
where $e_m$ is the base-complex edge in $E_{M}$ of $b_i$, $e_{lr}$ is the base-complex edge to be collapsed, and $e_{lr}^{\prime}$ is the created base-complex edge.

According to the energy function $E(m)$, some analysis on the structure of base-complex sheets can be performed.  The base-complex sheet has an interesting
property: the interior edges which are topology parallel to the dual face of sheet are all regular, the singular edges only exist in $E_M$ or $F_{L}$ and $F_{R}$, and the collapsing will introduce edges with a different valence. Hence, we can accurately predict the influence of collapsing.

\begin{figure}[t]
\setlength{\abovecaptionskip}{0.1cm}
\setlength{\belowcaptionskip}{-0.2cm}
  \centering
  \subfloat[]{
    \includegraphics[width=0.36\linewidth]{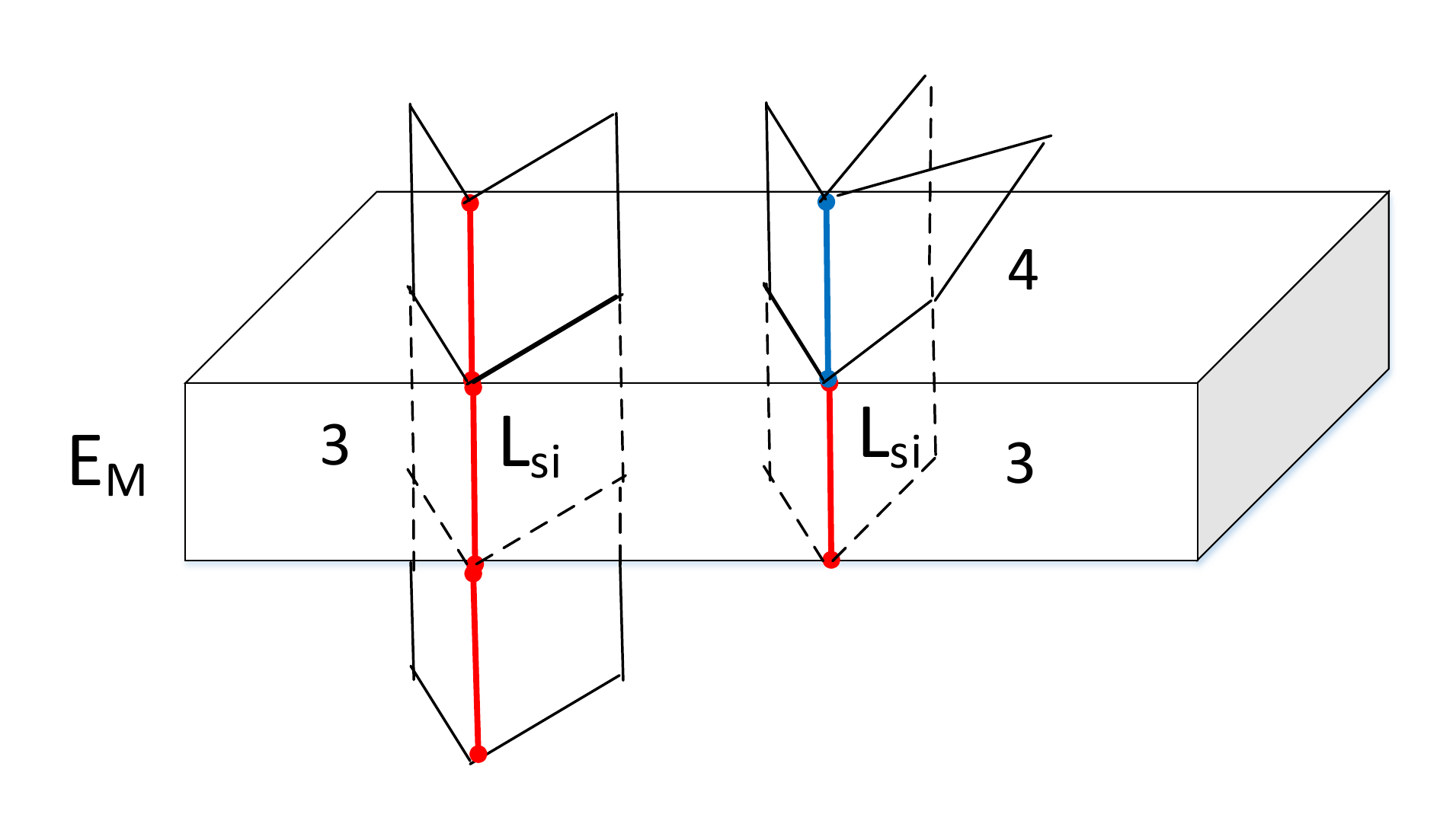}}
    \label{5a}
    \subfloat[]{
    \includegraphics[width=0.50\linewidth]{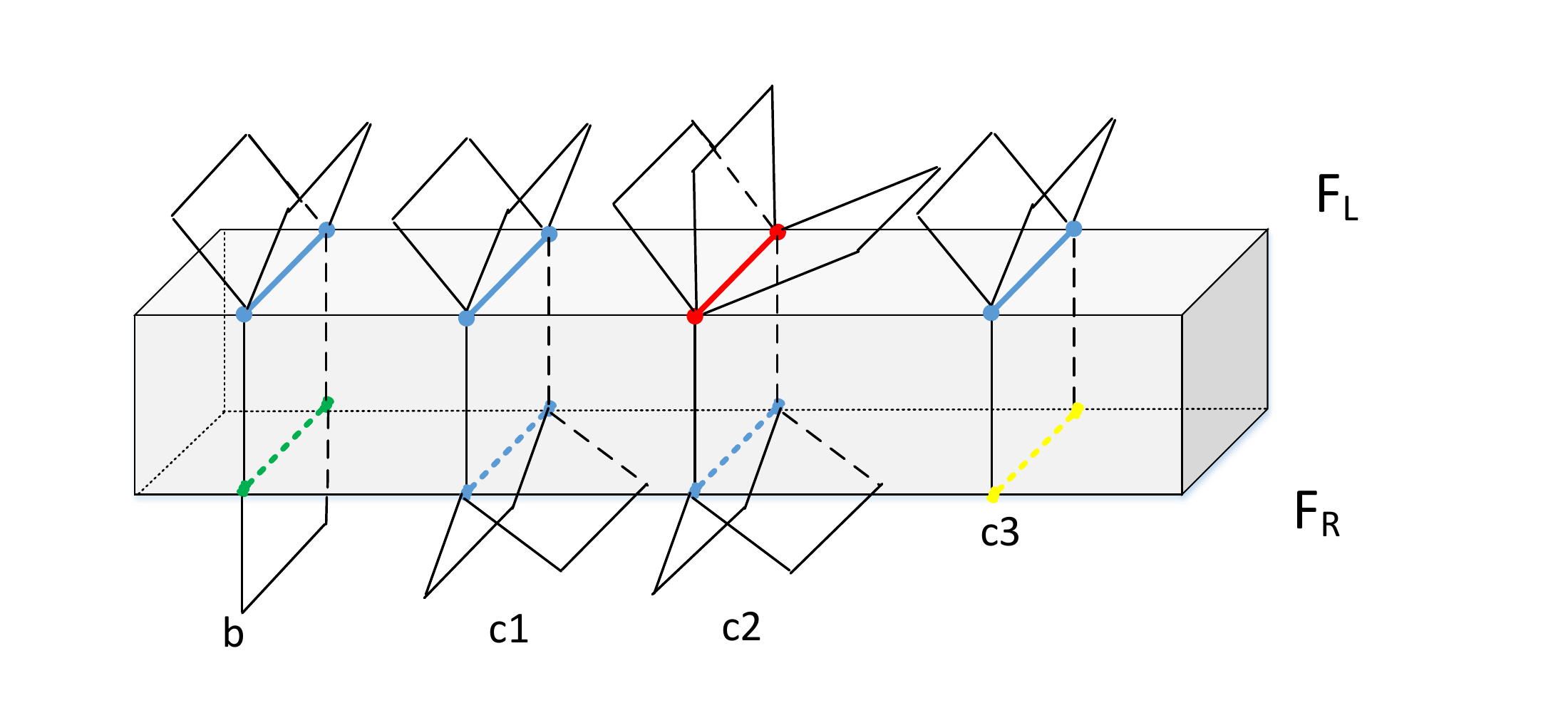}}
    \label{5b}
    \caption{Distribution of singularities in $E_M$, $F_L$ and $F_R$ is shown. The lines marked in green and black are regular edges, and all the other edges are singular. Two types of middle edges in $E_M$ are shown in (a), and four types of edge pairs on both sides are shown in (b).}
    \label{fig4}
\end{figure}

During a collapsing operation, the edges in the middle part will be eliminated. For a singular edge $L_{si}$, if the whole edge is contained in $E_M$ as $L_{mid}$, then the value of $E(m)$ will decrease.  This type of elimination is equivalent to creating new regular edges while collapsing. Moreover, the singularity structure will not change when $L_{mid}$ is a part of a singular edge $L_{si}$, and the type of collapsing does not affect the other part of the singular edge and the base-complex faces extended from it.  Such base-complex edges will not be considered in our valence calculation. Two types of $L_{si}$ are shown in Fig. \ref{fig4}(a).

Since a singular edge is completely contained in $F_{L}$ or $F_{R}$ of one or more base-complex sheets, the collapsing may remove the singular edges in both sides directly. Concerning the valence variation of edges in an edge pair of $F_{L}$ and $F_{R}$,  we have the following three cases which correspond to c1, c2 and c3 in Fig. \ref{fig4}(b) respectively: (c1) all the edges in $F_{L}$ and $F_{R}$ are  regular, (c2) edges in only one side of $F_{L}$ or $F_{R}$ are irregular, (c3) both edges in $F_L$ and $F_R$ are irregular. In case of (c1), the valence of the created edge will be regular; in case of (c2), the created edge will have the same valence as an irregular edge, and it does not affect the surrounding singularity configurations; in case of (c3), the valence of created edges will change, which means that the singularities of the rest part of hex-mesh will be changed, and  the flow direction of neighboring base-complex sheets might lead to different directions. Moreover, there are several configurations in case of (c3), the created edges might have different valences compared with base-complex edge pairs in $F_L$ and $F_R$. The singular structure will be simplified when the valence difference between irregular and regular edges decreases. In contrast, the removal making the valence of singular edges higher should be avoided. The created edges might not unknot self-interested sheets which are hard to remove, and it will greatly influence the final component reduction ratio, and cause an early termination for simplification.

To improve the convergence rate of $E(m)$, we greedily select the base-complex sheet which can effectively reduce $E(m)$ locally without introducing  edges with higher valence. The valence term is defined as
\begin{equation}
\begin{split}
E_{sv} = [DM - \beta (\sum\limits_{i}T ( K_{i}^{max} - K_{i}^{new} ) + \sum\limits_i K^m_i) ]/ DM,\
\end{split}
\end{equation}
in which
\begin{equation*}
\left\{\begin{split}
&K_{i}^{new} = \left| v(e_{i}^{new}) - q(e_{i}^{new}) \right|,\\
&K_i^m = \left| v(e_{mi}) - q(e_{mi}) \right|, \quad\quad\quad\quad\quad\quad
T(k) = \left\{
\begin{array}{rcl}
0.5k,    &   &k<0 \\
1.0,     &   &k=0 \\
k,       &   &k>0
\end{array} \right.\\
&K_{i}^{max} =  \max\left( \left| v(e_{li}) - q(e_{li}) \right| , \left| v(e_{ri}) - q(e_{ri}) \right| \right),
\end{split}\right.
\end{equation*}
where $e_{li}$ and $e_{ri}$ form an edge pair, and they belong to $F_L$ and $F_R$ respectively, $e_{mi}$ is the whole singular edge in $E_M$, $DM$ is a large value to control the scale of this term, which is set as the maximum number of $E_M$ in the hex-mesh. In our experiments, $\beta$ is set to be 1.67. For the purpose of minimizing the energy function,
the convergence rate will be faster when the value of $\beta (\sum\limits_{i = 0}^{n} T ( K_{i}^{max} - K_{i}^{new})+\sum\limits_i K^m_i))$ is much larger. $E_{sv}$ is a ranking term that encourages the collapsing candidate  which could eliminate more singularities.

\indent\textbf{Distortion term}. The distortion term $E_{sq}$ is an optional term for hex-mesh with complex structure, where the sheet passing through the regions with dense singularities often contains patches with serious distortion. Removing these
sheets can greatly improve the average value of Jacobians, and lead to a significant complexity reduction in simplification. Here we use the shape metric $f_{shape}$ of hexahedron \cite{knupp2003algebraic}  to measure the
sheet distortion. $f_{shape} = 1$ if the hexahedron is a cube with parallel faces, and $f_{shape} = 0$ if the hexahedron is degenerated, and
$f_{shape}$ is a scale-invariant. In our paper, we obtain the central difference of $f_{shape}$ in each element for three parametric directions, and select the maximum difference as the differential value of the hexahedron. From the experiments, we find that serious distortion happens when the differential value is up to $0.55$.
In this term, we use the central difference of $f_{shape}$ to find the regions with distortion, and twist is more serious while the differential value is bigger. Since the local parameterization can improve the element quality, removing regions with serious distortion in advance will increase the  average value of Jacobians locally.
 $E_{sq}$ is defined as
\begin{equation}
E_{sq}(s) = \ln (\sum\limits_{i=1}^{n} f_{i} + e)^{-1},
\end{equation}
\begin{equation*}
f_i = \left\{
\begin{array}{rcl}
0,           &   &d(i)<0.55 \\
d(i),     &   &d(i)\ge 0.55\\
\end{array} \right.  ,   \quad  d(i) = \max\limits_{0\le j \le 2}\left | f_{shape}(i+1,j)+f_{shape}(i-1,j)-2f_{shape}(i) \right |
\end{equation*}
where $f_{shape}(i,j)$ is of the neighboring element for the $i$-th element in the $j$-th parametric direction.

\indent\textbf{Width term}. The width term $E_{sd}$ in the weighted ranking function measures the width of sheet, which prevents wrong collapsing since if the sheet is too wide and then the collapsing will lead to big distortion on the boundary geometry and affect the adjacent sheets seriously. Hence it is reasonable to remove sheets with thin shape.  For this term, we use the width of base-complex edges in $E_M$, which is more accurate than the length between the vertex pair on surface.
In our framework, $E_{sd}$ is defined by combining the average width and the minimum length as follows,
\begin{equation}
E_{sd} = \left[ \left( \alpha_{a} \min\limits_{(v_l,v_r) \in P_V } d(v_l,v_r) + \alpha_{b}\bar{d}(v_l,v_r) \right)/\bar{L} \right]^{1/3}
\end{equation}
in which  $\bar{L}$ is the average length of element edges, $d(v_l,v_r)$ is the length of the base-complex edge connecting $v_l$ and $v_r$, and the weights $\alpha_a = 0.7$ and $\alpha_b = 0.3$.

\begin{figure}[t]
\setlength{\abovecaptionskip}{0.0cm}
\setlength{\belowcaptionskip}{-0.2cm}
  \centering
    \includegraphics[width=0.5\linewidth, height=0.28\linewidth]{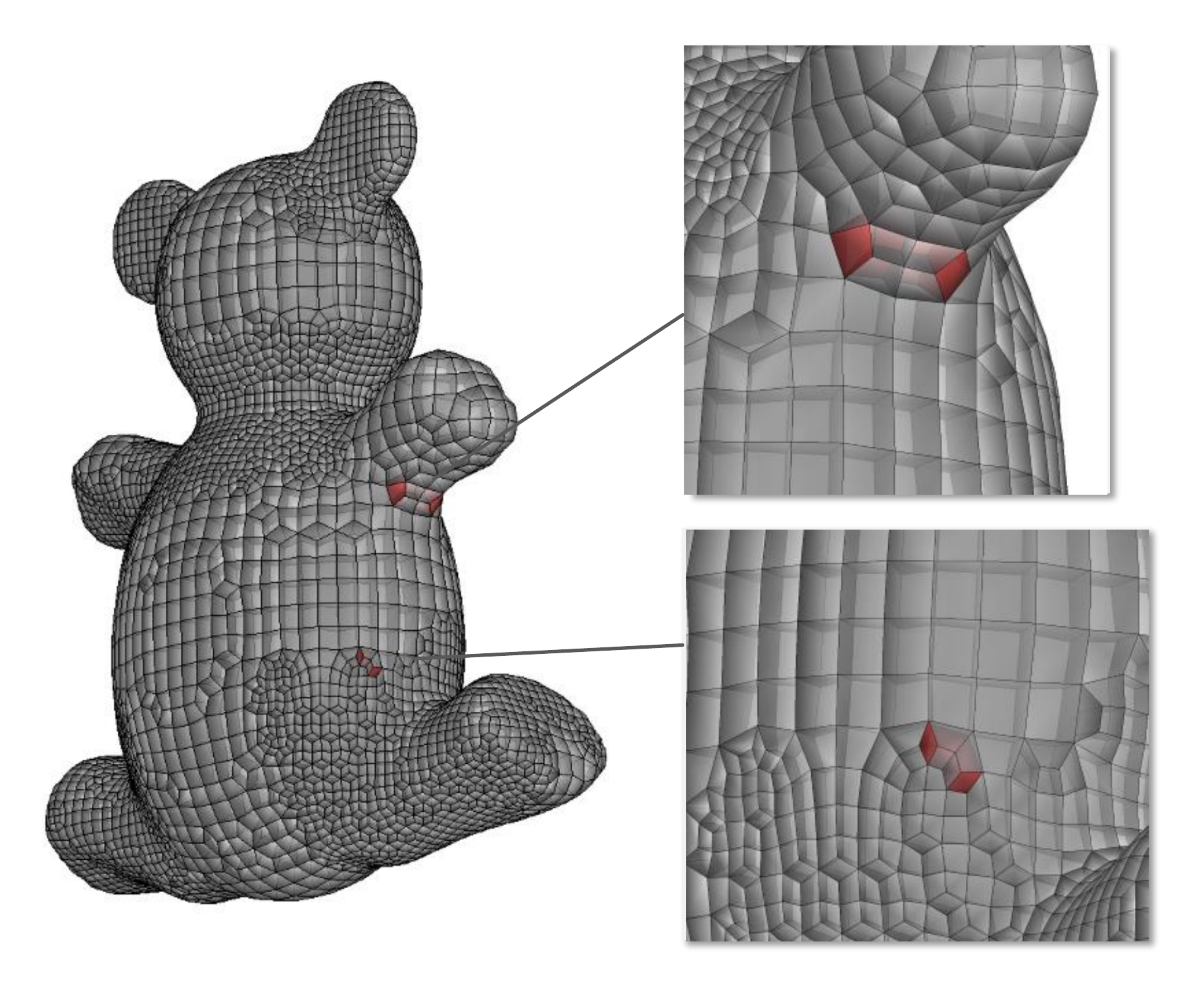}
    \caption{Two base-complex chords (red) in a toy mesh. The first chord is located in a patch near the feature edges (top right), and the second chord is located in the flat region (bottom right). The elimination of the first chord will lead to a significant  boundary geometry error. The proposed geometric error term can prevent this kind of collapsing effectively.}
    \label{fig5}
\end{figure}

\subsection{Ranking approach for base-complex chord}

The base-complex chord collapsing only influences one column of components, which is used to adjust regions with many edge pairs having a valence of 3$\sim$5. From our observation, edge pair with a valence of 3$\sim$5 often exists in the entangled sheets, which is difficult to eliminate. In order to untangle them, we propose a priority metric $E_c(c)$,
\begin{equation}
E_c(c) = k_{sq}(1-e^{-E_{cq}(c)}) + k_{sv}(1-e^{-E_{cv}(c)})
\end{equation}
in which $E_{cv}$ is the valence term and $E_{cq}$ is the geometry
error term.

\indent\textbf{Geometry error term}. The chord collapsing operation often leads to simplification results with inverted elements. We propose a simple strategy for priority processing on chords with narrow shape and smaller length.  The aspect ratio of a chord is defined as
the ratio of the average length of the main diagonal to the sub-diagonal, which is applied to the measurement of thickness. To reduce the collapsing effect on	
boundary geometry,  Gaussian curvature \cite{xu2006convergence} is used to measure the shape error locally after collapsing. In our implementation,  we use the variance of curvature to find patches with significant curvature changes.  A patch may contain sharp
features when its  variance of curvature is large as shown  in Fig. \ref{fig5}. The geometry error term $E_{cq}(c)$ is defined as
\begin{equation}
E_{cq}(c) = \frac{\bar{L} L_1(c)}{L_2(c)} \sqrt{\frac{\sum\limits_{i=1}^{N_v}(Q_{gi}-\bar{Q_g})^{2}}{N_v-1}}
\end{equation}
in which $L_1, L_2$ are the average length of the main diagonal and the sub-diagonal respectively, and $Q_{gi}$ is the Gaussian curvature of a vertex on two sides.

\indent\textbf{Valence error term}. The valence error term measures the valence error of four topological parallel edges. To eliminate entangled sheets and simplify the local complexity, we require that the three topological parallel edges created by collapsing should be all regular.
The ideal situation is that the valence error tends to be zero. In our framework, the valence error is set  as one of the optimization goals,
\begin{equation}
E_{cv}(c) = \beta D(c)/3N_{b}(c).
\end{equation}
In this step,  the edges with high valence will not be introduced, hence the candidates will not be pushed to the priority queue when $D(c)/3N_{b}(c)>0.9$.

\section{Sheet refinement}
\label{sec:refinement}

Sheet refinement is performed during the simplification pipeline in order to maintain the input mesh geometry with a similar number of elements to the user-defined target number. A similar method in \cite{gao2017robust} can be used to split one element on a specific sheet into two elements along the direction perpendicular to the parallel edges. In this paper, we propose an adaptive sheet refinement method to improve the accuracy of boundary geometry approximation.

In our implementation, we find that  choosing a sheet with the maximum width to refine is not a robust strategy, where some boundary patches with large boundary approximation error may not be refined. In our method, we firstly obtain the average length of all edges along the collapsing direction, and then compute the average Hausdorff distance ratio $HR(s)$ by the means of  point sampling for each sheet in the priority queue.  According to the descending order of $HR(s)$, the first four base-complex sheets will be selected in advance, and the average length in the collapsing direction is denoted as $\bar{L_b}$. We choose one from the first four sheets to perform refinement if $L_{b}>1.2\bar{L}$; otherwise, we refine the candidate with the maximum $\bar{L_b}$ and meeting the above condition. During simplification, collapsing operations may fail frequently due to the element quality and shape error constrains. In order to relax these constrains, we also perform the refinement process when a sheet collapsing fails. The base-complex sheets sharing $F_L$ and $F_R$ with the removed sheet are selected as candidates. The refinement process narrows the parameterized region of failed sheets, such that it reduces the shape error by introducing more elements, and the sheet may be collapsed in the next iteration. In addition, another criterion is introduced to control the number of elements strictly. For the input hex-mesh with $C_0$ elements, if the target number is $C_n$ before performing refinement, we check whether the number of hexahedra contained in a sheet is less than $1.5\times(C_0-C_n)$. This criterion can effectively prevent some sheets being refined repeatedly.

\begin{figure*}[t]
\setlength{\abovecaptionskip}{0.1cm}
\setlength{\belowcaptionskip}{-0.2cm}
\centering
    \subfloat[]{
        \includegraphics[width=0.95\linewidth]{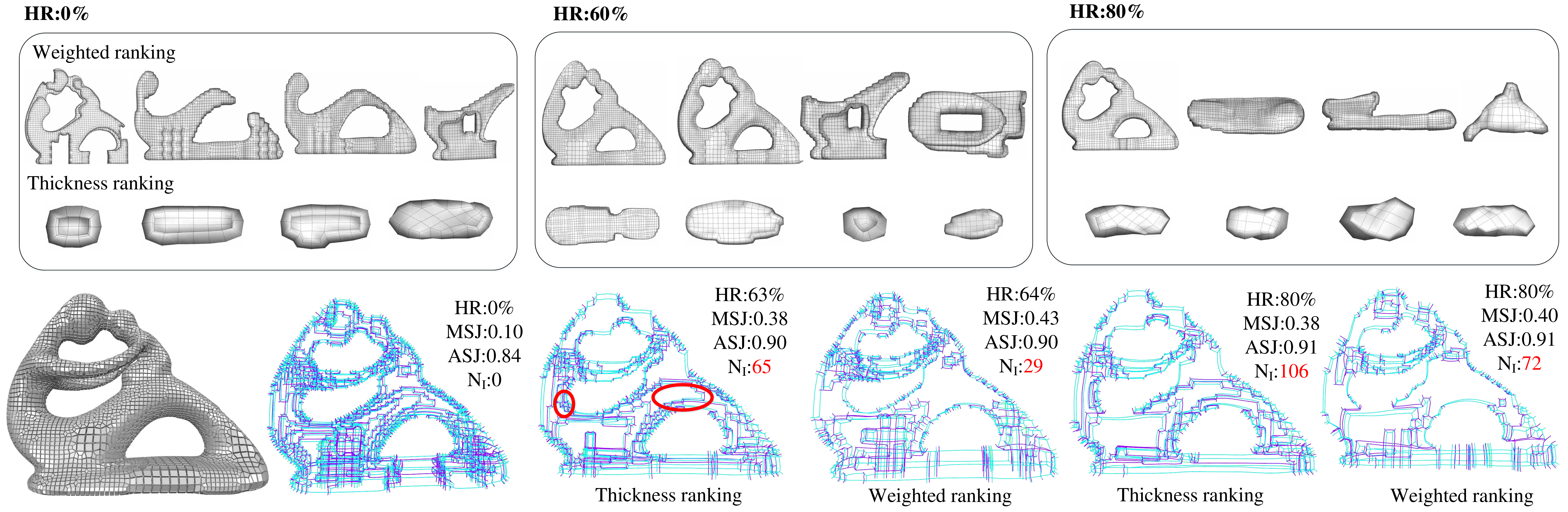}}\\
    \subfloat[]{
        \includegraphics[width=0.55\linewidth]{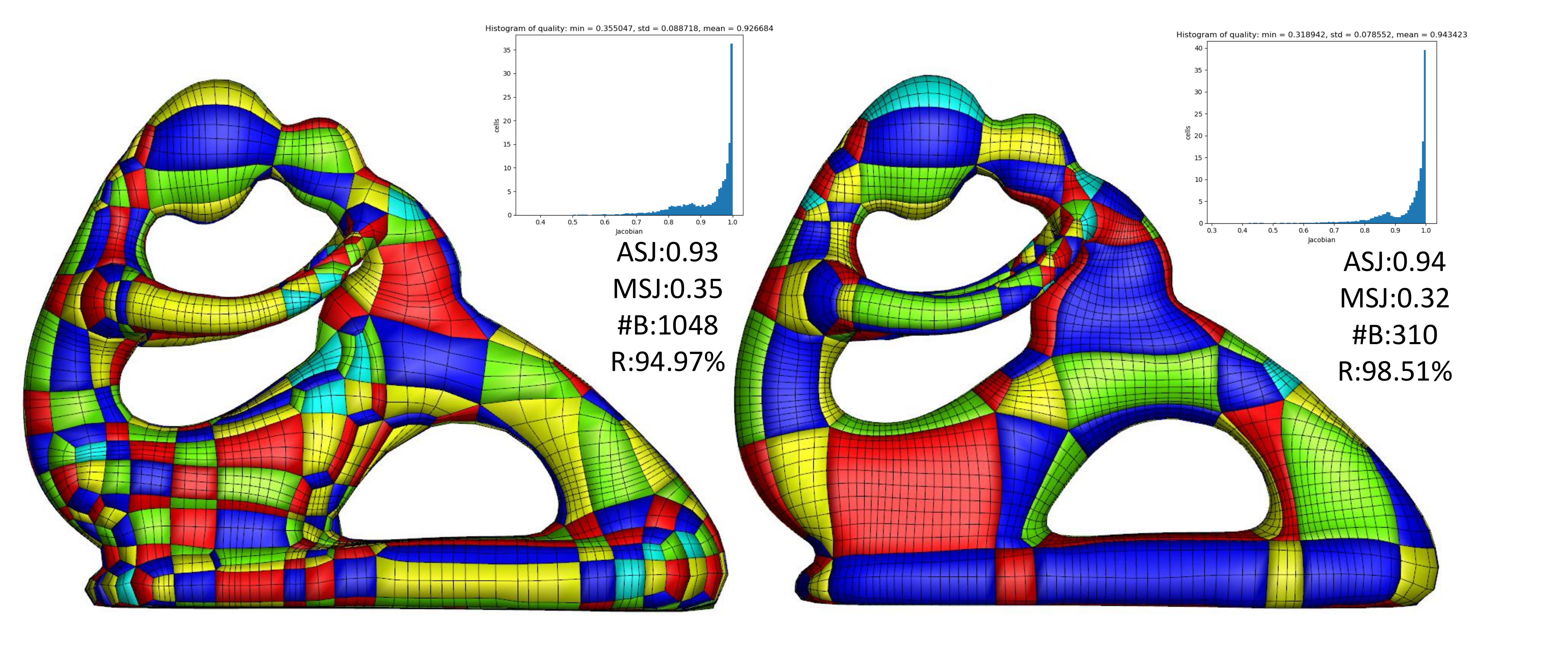}}
    \subfloat[]{
        \includegraphics[width=0.37\linewidth]{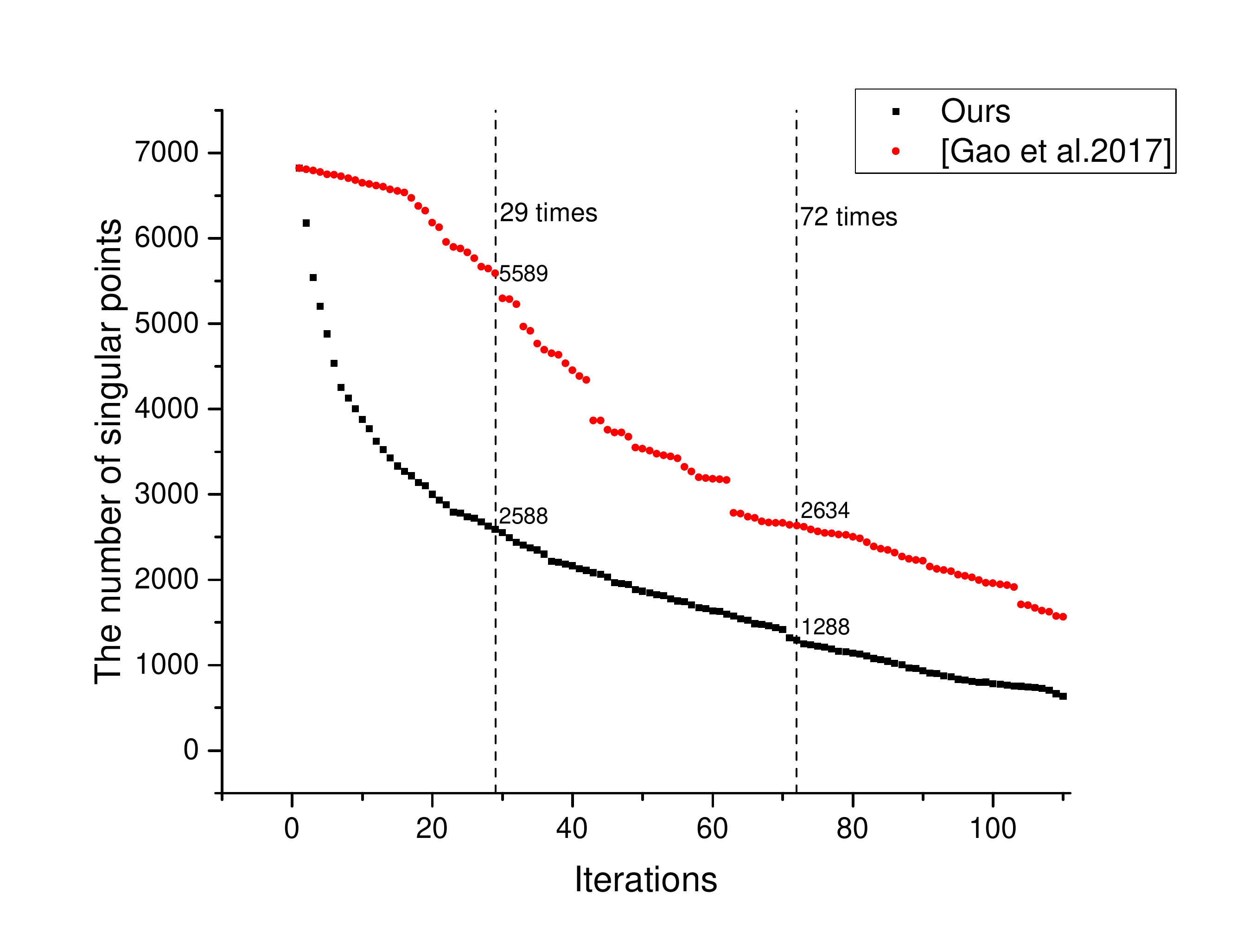}}
\caption{Simplification results of the fertility mesh with different complexity reductions, including our weighted ranking approach and the thickness ranking method \cite{gao2017robust} are shown in (a). Our ranking method can effectively decrease the iteration steps ($N_I$) and improve the simplification results around regions with dense singularities as shown in the singular structure highlighted with red circles. The top 4 candidates in each sequence are also shown when the simplification rates achieve $0\%$, $60\%$ and $80\%$. The simplification results are shown in (b), and the statistics of iterations are shown in (c).} \label{fig6}
\end{figure*}

\begin{figure}[!htb]
\setlength{\abovecaptionskip}{-0.3cm}
\setlength{\belowcaptionskip}{-0.5cm}
    \centering
        \includegraphics[width=1.0\linewidth]{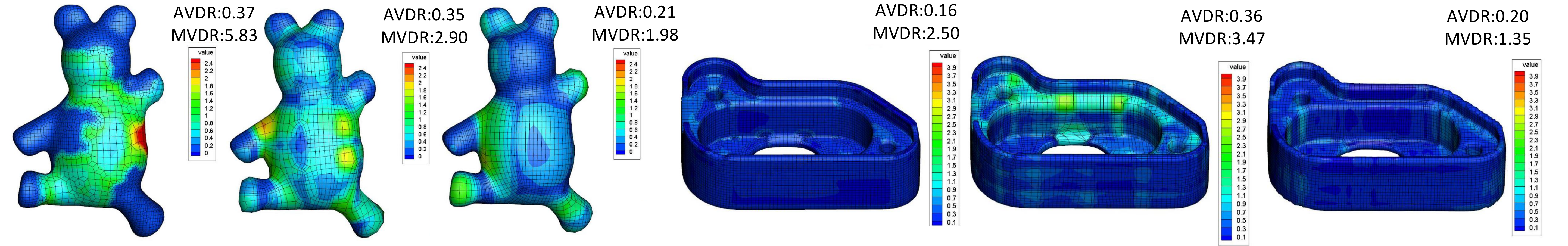}
        \caption{Simplification results of toy2 and lock.  From left to right, the input meshes, results of thickness ranking \cite{gao2017robust} and our weighted ranking results are shown respectively. The color mapping shows the value of VDR, which illustrates that our weighted ranking method can achieve a significant improvement on uniformity.}
        \label{fig7}
\end{figure}

\begin{figure*}[t]
\setlength{\abovecaptionskip}{0.0cm}
\setlength{\belowcaptionskip}{-0.2cm}
  \centering
    \includegraphics[width=0.95\linewidth]{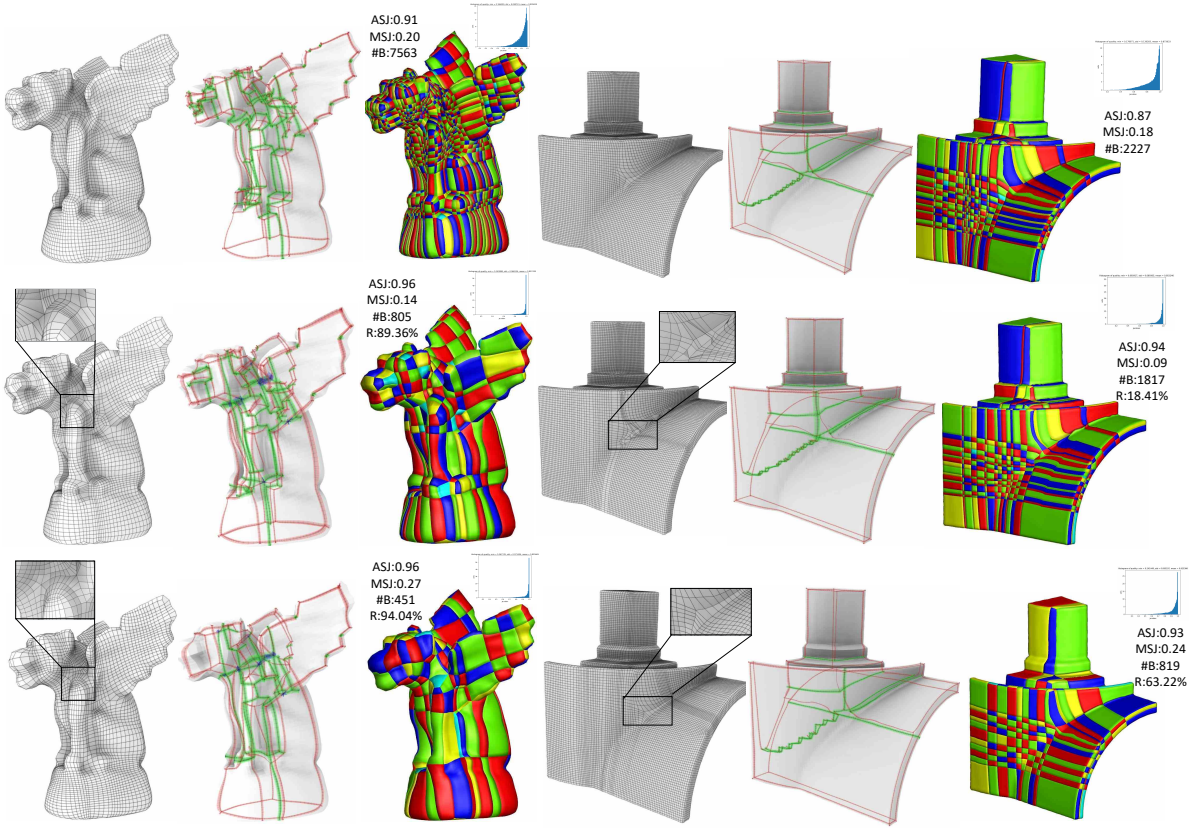}
\caption{Simplification results on meshes generated by the polycube-based method, gargoyle (left) mesh is generated by \cite{Huang2014ℓ}, and the stab (right) is generated by \cite{Gregson2011All}. From top to bottom,  the input hex-mesh, simplification results of thickness ranking \cite{gao2017robust} and our weighted ranking results are presented. For each example, we show the information of scaled Jacobian, singularity structure, and the base-complex components with different colors. } \label{fig8}
\end{figure*}

\begin{figure*}[htb]
\setlength{\abovecaptionskip}{0.0cm}
\setlength{\belowcaptionskip}{-0.2cm}
  \centering
    \includegraphics[width=1.05\linewidth]{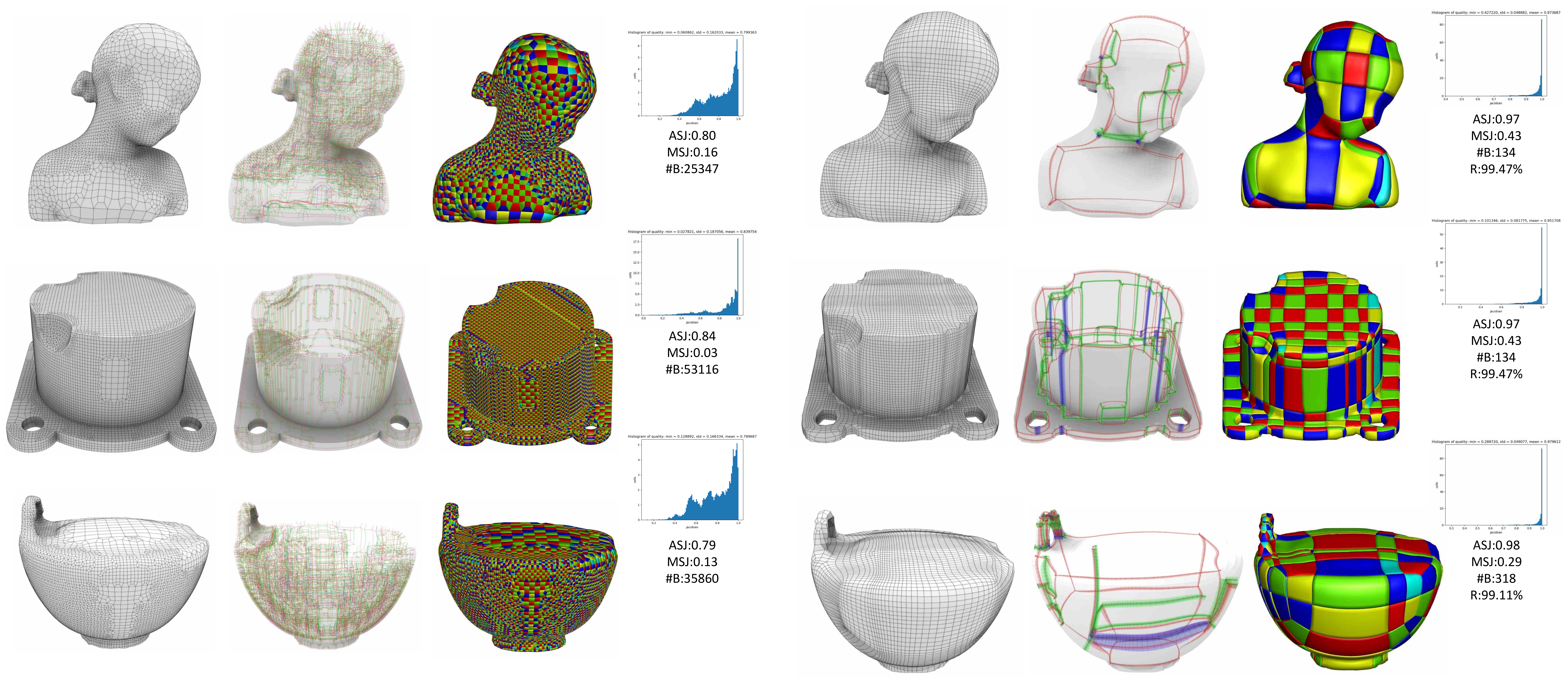}
\caption{The simplification results on meshes generated by octree-based methods, including the bimbia, deckel and bottle models.  From left to right,  the input mesh, singularity structure (the singular edges with a valence of 5 marked in green, and a valence of 3 marked in red, the valence of edges with other colors is $>$5),  base-complex of original meshes and simplified results. } \label{fig9}
\end{figure*}

\begin{figure*}[!htb]
\setlength{\abovecaptionskip}{-0.2cm}
  \centering
    \includegraphics[width=1.02\linewidth]{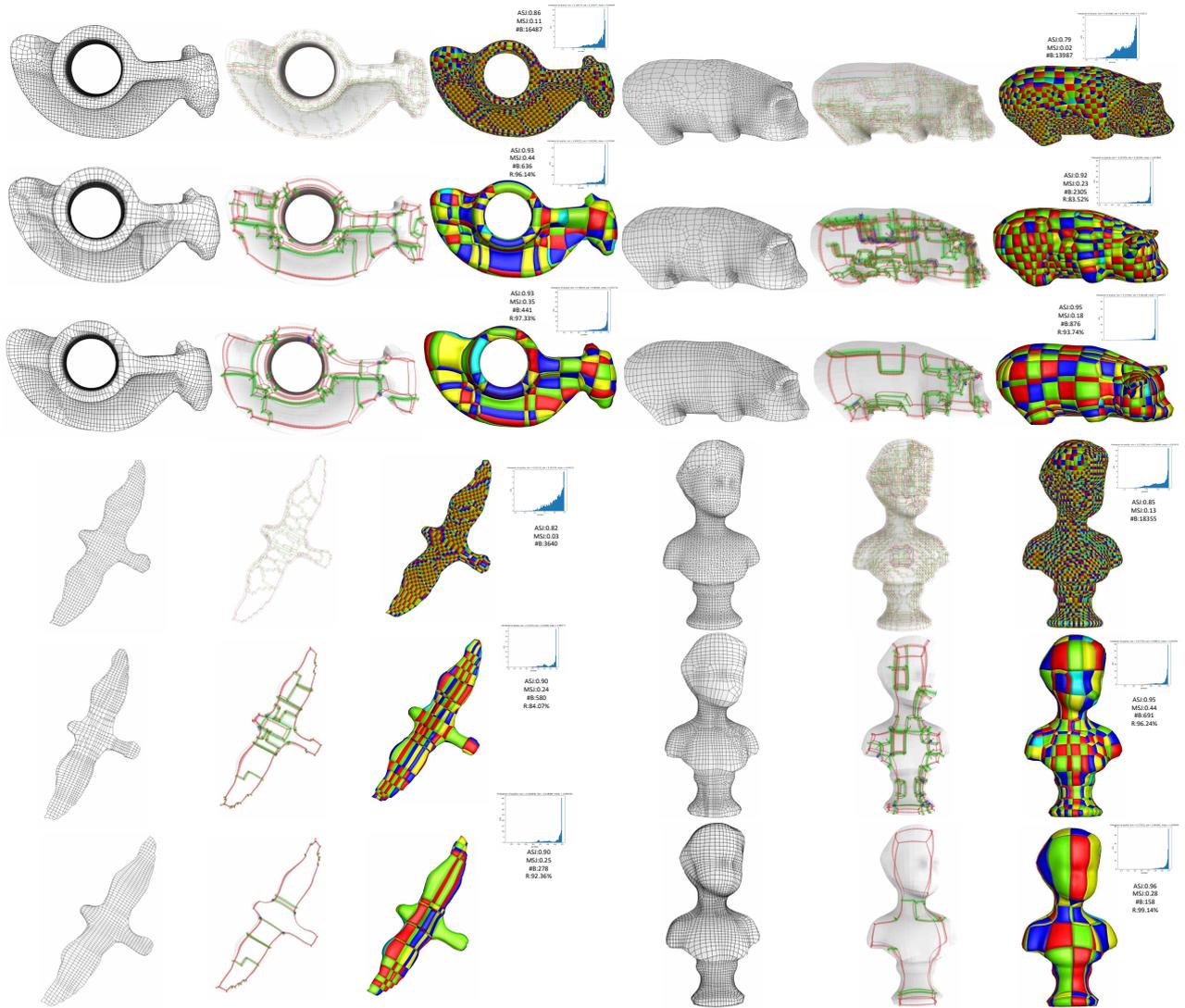}
\caption{From left to right,  the input octree-based hex-mesh, simplification results of \cite{gao2017robust} and our results. For each example, we show the scaled Jacobian, singularity structure,  and base-complex components with different colors.} \label{fig10}
\end{figure*}

\begin{figure*}[!htb]
\setlength{\abovecaptionskip}{0.0cm}
\setlength{\belowcaptionskip}{-0.2cm}
  \centering
    \includegraphics[width=0.72\linewidth]{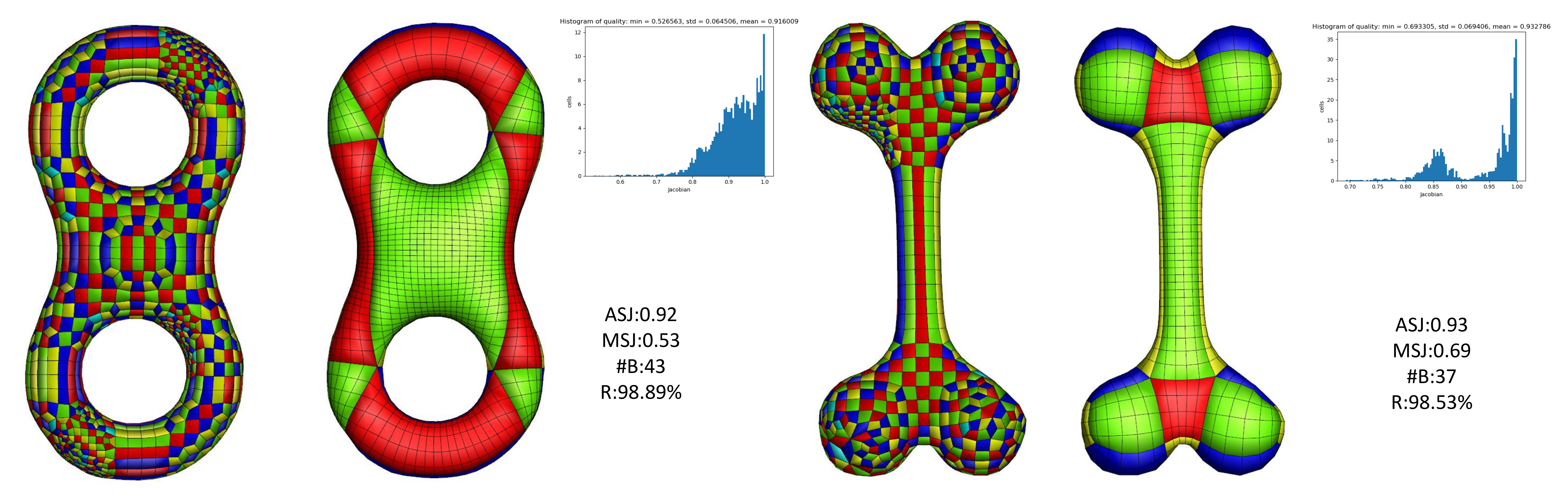}
    \caption{Simplified results on octree-based meshes \cite{meshgems2015volume}, and their singularity structures are similar to polycube-based meshes.} \label{fig11}
\end{figure*}

\section{Experimental results}
\label{sec:example}

We tested our algorithm on a four-core i7 processor with 8 GB memory. The maximal number of iterations of the SLIM solver is set as 5, and we set $r_h = 1\%$ (the threshold of Hausdorff distance ratio that defined by the user, the simplification rate becomes larger when $r_h$ increases) and $r_{|H|} = 1.0$ (the rate of the target number over the number of elements in the input mesh) for all experiments and figures. We also report the number of hex elements ($\#H$), the number of base-complex components ($\#BC$) and the minimal, average and standard variance value of scaled Jacobians (MSJ/ASJ/Std). The boundary geometry error is measured by the Hausdorff distance ratio (HR). For the experiments on the database given in \cite{gao2017robust}, we perform the proposed method in $65\%$ of meshes in this database. Most meshes achieve higher simplification ratio compared with \cite{gao2015hexahedral} and \cite{gao2017robust}, and the average simplification rate for these meshes is $88\%$.


\indent\textbf{Weighted ranking candidates}. Here we show some comparison results of the thickness ranking method \cite{gao2017robust} and the proposed weighted ranking method.
In Fig. \ref{fig6}(a), we show the top 4 candidates in the fertility mesh when the simplification rates achieve $0\%$, $60\%$ and $80\%$ respectively. In the initial priority queue, our weighted ranking term can effectively pick up the base-complex sheets with serious distortion and close-loop configurations. Moreover, the number of singularities can also be reduced faster.  For a simplification rate of $60\%$,
the thickness ranking method needs $65$ iterations, and our proposed method only needs $28$ iterations.  For the comparison results as shown in Fig. \ref{fig6}(a), when the simplification rates reach $60\%$ and $80\%$, our ranking algorithm  can preferentially remove sheets to promote singular edge elimination, and the regions with dense singularities (marked with red circles) have been greatly improved. Compared with the simplification results by thickness ranking, regions with dense singular edges can be successfully eliminated by our method, and self-intersected sheets can be removed as well at the same time. In the simplification process, the distortion term is used to eliminate elements with poor shape quality, and to spread
the distortion to neighboring elements while gradually improving the value of MSJ/ASJ in the hex-mesh. Our ASJ is better than thickness ranking during these three stages,
and we can achieve $12.66\%$ ASJ improvement over the input and $2.20\%$ ASJ improvements over the simplification result by \cite{gao2017robust}. The average running time of the entire dataset is $71$ minutes, which is slightly slower than \cite{gao2017robust}.


\indent\textbf{Element uniformity}. In the proposed approach,  we use local parameterization  to improve the uniformity of hex-mesh elements. We also propose a measurement of element uniformity called the \emph{volume deviation ratio} (VDR), which is denoted as the standard  volume deviation of neighboring elements divided by the average element volume. The range of VDR is $(0,\infty]$, and the uniformity is better while the value is closer to 0 (for all elements with the same volume, VDR$=0$). Compared with the thickness ranking method \cite{gao2017robust}, our simplification results have $30.17\%$ and $7.04\%$ improvement in the average volume deviation ratio (AVDR) and the max volume deviation ratio (MVDR).  In our experiments, the average AVDR and MVDR of meshes from polycube-base methods are $0.19$ and $2.78$ respectively, and the average AVDR/MVDR are $0.25/2.54$  in the simplification results of octree-base meshes.  AVDR and MVDR gain $35.56\%$ and $10.86\%$ improvement compared with the thickness ranking approach for octree-based meshes. Two comparison examples are  shown in Fig. \ref{fig7} with the VDR colormap.

\indent\textbf{Simplification of hex-mesh from polycube-based methods}.    For hex-meshes generated by polycube-based methods  \cite{Gregson2011All,Huang2014ℓ,Fang2016All},  the singularity structures are completely distributed on the surface, and the distribution of singular edges is sparse. Hence, the valence term has a small effect, and the weights $k_{sd}$ and $k_{sv}$ are set to a smaller value ($k_{sd}=0.7$ and $k_{sv}=0.3$  in our experiments). As shown in Fig. \ref{fig8} and Table \ref{table2}, the proposed approach can  achieve a higher base-complex component reduction with similar element quality as the results in \cite{gao2017robust}. In our experiments,  the average scaled Jacobian is improved to  $0.95$, and the meshes obtain $30.17\% / 7.04\%$ improvement for  AVDR and MVDR  compared with \cite{gao2017robust}, respectively. Moreover, the  average components reduction ratio is promoted to $71.51\%$, and some results are close to the structure of meshes generated by \cite{Gianmarco2016Polycube}.

\begin{table}[]
\setlength{\abovecaptionskip}{0.1cm}
\setlength{\belowcaptionskip}{0.1cm}
\caption{Statistics of meshes generated by octree-based methods.}
\sffamily
\resizebox{\textwidth}{18mm}{
\begin{tabular}{l|r|r|r|r|r|r|r|r|r|r|r|r|r}
\hline
\multicolumn{1}{c|}{\textbf{}} & \multicolumn{5}{c|}{\textbf{Input hex mesh}}                                                                                                                                                      & \multicolumn{8}{c}{\textbf{Simplified result}}                                                                                                                                                                                                                                                                             \\ \hline
\textbf{Model}                 & \multicolumn{1}{l|}{\textbf{\#H}} & \multicolumn{1}{l|}{\textbf{\#BC}} & \multicolumn{1}{l|}{\textbf{MSJ}} & \multicolumn{1}{l|}{\textbf{ASJ}} & \multicolumn{1}{l|}{\textbf{Std}} & \multicolumn{1}{l|}{\textbf{\#H}} & \multicolumn{1}{l|}{\textbf{\#BC}} & \multicolumn{1}{l|}{\textbf{MSJ}} & \multicolumn{1}{l|}{\textbf{ASJ}} & \multicolumn{1}{l|}{\textbf{Std}} & \multicolumn{1}{l|}{\textbf{HR (\%)}} & \multicolumn{1}{l|}{\textbf{R (\%)}} & \multicolumn{1}{l}{\textbf{Time (m)}} \\ \hline
Bimba (Fig.9)                   & 25,347                            & 25,347                             & 0.06                              & 0.80                              & 0.162                                   & 27,900                            & 134                                & 0.43                              & 0.97                              & 0.049                                   & 0.95                                 & 99.47                               & 103.48                                   \\
Bottle (Fig.9)                  & 35,886                            & 35,860                             & 0.13                              & 0.79                              & 0.167                                   & 34,558                            & 266                                & 0.22                              & 0.98                              & 0.054                                   & 0.91                                 & 99.26                               & 200.97                                   \\
Deckle (Fig.9)                  & 53,658                            & 53,116                             & 0.03                              & 0.84                              & 0.187                                   & 53,680                            & 806                                & 0.10                              & 0.95                              & 0.082                                   & 1.00                                 & 98.48                               & 793.93                                   \\
Fertility (Fig.6)               & 21,370                            & 20,840                             & 0.10                              & 0.84                              & 0.150                                   & 21,016                            & 310                                & 0.32                              & 0.94                              & 0.079                                   & 0.87                                 & 98.51                               & 153.83                                   \\
Toy1 (Fig.5)                    & 18947                             & 18883                              & 0.12                              & 0.81                              & 0.161                                   & 15784                             & 144                                & 0.51                              & 0.96                              & 0.059                                   & 0.66                                 & 99.23                               & 48.53                                    \\
Toy2 (Fig.7)                    & 14,288                            & 14,288                             & 0.15                              & 0.81                              & 0.158                                   & 13,952                            & 129                                & 0.49                              & 0.96                              & 0.059                                   & 0.90                                 & 99.10                               & 48.59                                    \\
Lock (Fig.7)                    & 28,753                            & 25,720                             & 0.01                              & 0.80                              & 0.244                                   & 28,501                            & 2,990                              & 0.17                              & 0.93                              & 0.109                                   & 0.91                                 & 88.37                               & 381.46                                   \\
Eight (Fig.11)                  & 4,571                             & 3,867                              & 0.17                              & 0.78                              & 0.155                                   & 5,428                             & 43                                 & 0.53                              & 0.92                              & 0.065                                   & 0.69                                 & 98.89                               & 7.53                                     \\
Bone (Fig.11)                   & 2,751                             & 2,520                              & 0.15                              & 0.78                              & 0.159                                   & 2,484                             & 37                                 & 0.69                              & 0.93                              & 0.069                                   & 0.75                                 & 98.53                               & 4.24                                     \\ \hline
\end{tabular}}
\label{table1}
\vspace{-12pt}
\end{table}

\indent\textbf{Simplification of hex-mesh from octree-based methods}. Octree-based hex-meshing approaches often generate a complex structure  with dense local singularities.  In \cite{gao2017robust}, the greedy collapsing by thickness ranking was utilized under a set of filters. It can not find a coarser structure in the hex-mesh with a large number of interior singularities and kinking, since the thickness ranking term does not have a direct effect on singularity removal. The corresponding simplification \cite{gao2017robust} has a slow convergence rate, and it  can achieve an average simplification rate around $86\%$ for the hex-mesh database.  Instead, our weighted ranking method can obtain a much simpler singularity structure with much fewer base-complex components. The average simplification rate in the proposed framework can increase $93.56\%$ with respect to the initial number of base-complex components in the input hex-mesh, and gain $7.40\%$ improvement compared with \cite{gao2017robust}. Moreover, in the proposed framework,  adaptive refinement is performed during the simplification process, which can effectively maintain the quality of boundary geometry and promote the simplification process under the constraint of $r_h$. Our ASJ/MSJ achieves $0.93/0.32$, and gain $14.02\%$ ASJ improvements over the thickness ranking method.  Some simplification  results are shown in Fig. \ref{fig9}, and statistics are presented in Table \ref{table1}.  Comparison examples with \cite{gao2017robust} are also presented in Fig. \ref{fig10} and Table \ref{table2}.

\begin{table}[]
\setlength{\abovecaptionskip}{0.1cm}
\setlength{\belowcaptionskip}{0.1cm}
\caption{Comparison with \cite{gao2017robust}. $\#H$ is the number of hex-elements, $\#BC$ is the number of base-complex components, $Std$ is the standard deviation of the scaled Jacobians, $HR$ stands for the Hausdorff distance, and $R$ is the simplification rate.}
\sffamily
\resizebox{\textwidth}{34mm}{
\begin{tabular}{clrrrrrrrrrr}
\hline
\multicolumn{2}{c}{\textbf{Model}}                                             & \textbf{\#H} & \textbf{\#BC}                & \textbf{MSJ} & \textbf{ASJ} & \textbf{Std} & \textbf{AVDR} & \textbf{MVDR} & \textbf{HR (\%)} & \textbf{R (\%)} & \textbf{Time (m)} \\ \hline
\multicolumn{1}{c|}{}                                  & Input                 & 21,167       & 25,669                       & 0.20         & 0.91         & 0.907             & 0.11          & 0.64          &                 &                &                      \\ \cline{2-12}
\multicolumn{1}{c|}{}                                  & Thickness ranking & 22,524       & {\color[HTML]{3531FF} 805}   & 0.14         & 0.96         & 0.068             & 0.30          & 3.52          & 0.98            & 89.36          & 30.67                \\ \cline{2-12}
\multicolumn{1}{c|}{\multirow{-3}{*}{Gargoyle (Fig.8)}} & Weighted ranking                  & 23,352       & {\color[HTML]{FE0000} 451}   & 0.27         & 0.96         & 0.071             & 0.20          & 3.47          & 0.92            & 94.04          & 41.22                \\ \hline
\multicolumn{1}{c|}{}                                  & Input                 & 84,489       & 2,227                        & 0.18         & 0.87         & 0.130             & 0.05          & 1.29          &                 &                &                      \\ \cline{2-12}
\multicolumn{1}{c|}{}                                  & Thickness ranking & 80,295       & {\color[HTML]{3531FF} 1,817} & 0.09         & 0.95         & 0.069             & 0.10          & 2.74          & 0.77            & 18.41          & 46.89                \\ \cline{2-12}
\multicolumn{1}{c|}{\multirow{-3}{*}{Stb (Fig.8)}}      & Weighted ranking                  & 83,678       & {\color[HTML]{FE0000} 819}   & 0.24         & 0.93         & 0.092             & 0.10          & 1.35          & 0.97            & 63.22          & 60.32                \\ \hline
\multicolumn{1}{c|}{}                                  & Input                 & 16,608       & 16,487                       & 0.11         & 0.86         & 0.139             & 0.18          & 2.69          &                 &                &                      \\ \cline{2-12}
\multicolumn{1}{c|}{}                                  & Thickness ranking & 10,278       & {\color[HTML]{3531FF} 636}   & 0.44         & 0.93         & 0.081             & 0.38          & 1.77          & 0.99            & 96.14          & 32.25                \\ \cline{2-12}
\multicolumn{1}{c|}{\multirow{-3}{*}{Rocker (Fig.10)}}  & Weighted ranking                  & 10,790       & {\color[HTML]{FE0000} 441}   & 0.35         & 0.93         & 0.088             & 0.26          & 1.75          & 0.99            & 97.33          & 50.55                \\ \hline
\multicolumn{1}{c|}{}                                  & Input                 & 13,987       & 13,987                       & 0.02         & 0.79         & 0.168             & 0.46          & 8.02          &                 &                &                      \\ \cline{2-12}
\multicolumn{1}{c|}{}                                  & Thickness ranking & 10,704       & {\color[HTML]{3531FF} 2,305} & 0.23         & 0.92         & 0.102             & 0.38          & 3.89          & 0.99            & 83.52          & 31.24                \\ \cline{2-12}
\multicolumn{1}{c|}{\multirow{-3}{*}{Pig (Fig.10)}}     & Weighted ranking                  & 11,218       & {\color[HTML]{FE0000} 876}   & 0.18         & 0.95         & 0.086             & 0.23          & 3.02          & 0.99            & 93.74          & 38.69                \\ \hline
\multicolumn{1}{c|}{}                                  & Input                 & 4,247        & 3,640                        & 0.03         & 0.82         & 0.159             & 0.18          & 0.51          &                 &                &                      \\ \cline{2-12}
\multicolumn{1}{c|}{}                                  & Thickness ranking & 2,868        & {\color[HTML]{3531FF} 580}   & 0.24         & 0.90         & 0.117             & 0.36          & 2.43          & 1.00            & 84.07          & 14.57                \\ \cline{2-12}
\multicolumn{1}{c|}{\multirow{-3}{*}{Bird (Fig.10)}}    & Weighted ranking                  & 2,935        & {\color[HTML]{FE0000} 278}   & 0.25         & 0.90         & 0.127             & 0.23          & 1.45          & 0.95            & 92.36          & 16.54                \\ \hline
\multicolumn{1}{c|}{}                                  & Input                 & 19,075       & 18,355                       & 0.13         & 0.85         & 0.151             & 0.28          & 3.20          &                 &                &                      \\ \cline{2-12}
\multicolumn{1}{c|}{}                                  & Thickness ranking  & 17,680       & {\color[HTML]{3531FF} 691}   & 0.44         & 0.95         & 0.070             & 0.33          & 4.36          & 0.98            & 96.24          & 113.67               \\ \cline{2-12}
\multicolumn{1}{c|}{\multirow{-3}{*}{Buste (Fig.10)}}   & Weighted ranking                  & 16,336       & {\color[HTML]{FE0000} 158}   & 0.28         & 0.96         & 0.065             & 0.23          & 2.82          & 0.97            & 99.14          & 53.55                \\ \hline
\end{tabular}}
\label{table2}
\vspace{-6pt}
 \end{table}

More importantly, octree-based meshes can be simplified into a similar singularity structure as polycube meshes. As shown in Fig. \ref{fig11}, singularities were mainly distributed on the boundary. The simplification rate achieves  $98\%$, and the interior singular edges are eliminated completely.

\section{Conclusion and future work}
\label{sec:conclude}

In this paper, an improved singularity structure simplification method of hex meshes is proposed  based on a weighted ranking function, which is a combination of the valence prediction function of local singularity structure, shape quality metric of elements and the width of base-complex sheets/chords. Local optimization and adaptive sheet refinement are also proposed to improve the element quality of simplified hex-mesh.  Compared with the thickness ranking method,  simpler singularity structure with fewer base-complex components can be achieved by the proposed weighted ranking approach while achieving better mesh quality and Hausdorff distance ratio. The proposed approach has a few limitations. Sharp features can not be preserved very well on the boundary, and the boundary approximation error may increase in models with high genus. Possible solution might be a more strictly feature edge extraction and vertex mapping. In the future, we will apply the proposed hex-mesh simplification method  to volume parameterization, which is a bottleneck in isogeometric analysis.




\nocite{*}
\bibliography{mybibfile}

\end{document}